\documentclass[iicol,sn-mathphys-num]{sn-jnl}
\usepackage[utf8]{inputenc}
\usepackage{subcaption}
\usepackage{overpic}
\usepackage{threeparttable}
\usepackage{textgreek}
\usepackage{booktabs}
\usepackage{multirow}
\usepackage{csquotes} 
\usepackage[version=4]{mhchem}
\newcommand{\cawo}{\ce{CaWO_4}}
\newcommand{\alo}{\ce{Al_2O_3}}
\newcommand{\lialo}{\ce{LiAlO_2}}
\usepackage[separate-uncertainty=true, free-standing-units=true]{siunitx}
\DeclareSIUnit\eVc{\eV\per\clight\squared}
\DeclareSIUnit\clight{\text{\!\ensuremath{c}}}
\usepackage{cleveref}
\crefname{lstlisting}{listing}{listings}
\Crefname{lstlisting}{Listing}{Listings}
\newcommand{\geant}{\textsc{Geant4}}
\newcommand{\cresst}{\textsc{Cresst}}
\newcommand{\impcresst}{Imp\textsc{Cresst}}
\newcommand{\cresstds}{\textsc{CresstDS}}
\usepackage{listings}
\usepackage{layouts}
\usepackage{subcaption}

\title{\impcresst{} - A versatile simulation tool focusing on solid-state detectors at keV energies}

\begin{document}
\author[1]{\fnm{G.} \sur{Angloher}}
\author*[2,3]{\fnm{S.} \sur{Banik}}\email{samir.banik@oeaw.ac.at}
\author[1,10]{\fnm{A.} \sur{Bento}}
\author[4]{\fnm{A.} \sur{Bertolini}}
\author*[5]{\fnm{R.} \sur{Breier}}\email{robert.breier@fmph.uniba.sk}
\author[6]{\fnm{C.} \sur{Bucci}}
\author[2]{\fnm{J.} \sur{Burkhart}}
\author[4]{\fnm{L.} \sur{Burmeister}}
\author[1]{\fnm{L.} \sur{Canonica}}
\author[1]{\fnm{F.} \sur{Casadei}}
\author[1]{\fnm{E.} \sur{Cipelli}}
\author[1, 6]{\fnm{S.} \sur{Di Lorenzo}}
\author[7]{\fnm{J.} \sur{Dohm}}
\author[1]{\fnm{F.} \sur{Dominsky}}
\author[8,11]{\fnm{A.} \sur{Erb}}
\author[4]{\fnm{E.} \sur{Fascione}}
\author[8]{\fnm{F.} \sur{von Feilitzsch}}
\author[2]{\fnm{S.} \sur{Fichtinger}}
\author[3,2]{\fnm{D.} \sur{Fuchs}}
\author[2,3]{\fnm{A.} \sur{Fuss}}
\author[2]{\fnm{V.M.} \sur{Ghete}}
\author[6]{\fnm{P.} \sur{Gorla}}
\author[1,12]{\fnm{P.V.} \sur{Guillaumon}}
\author[1]{\fnm{D.} \sur{Hauff}}
\author[5]{\fnm{M.} \sur{Ješkovsk\'y}}
\author[7]{\fnm{J.} \sur{Jochum}}
\author[8]{\fnm{M.} \sur{Kaznacheeva}}
\author*[2]{\fnm{H.} \sur{Kluck}}\email{holger.kluck@oeaw.ac.at}
\author[9]{\fnm{H.} \sur{Kraus}}
\author[4]{\fnm{B.} \sur{von Krosigk}}
\author[1]{\fnm{A.} \sur{Langenk\"amper}}
\author[1]{\fnm{M.} \sur{Mancuso}}
\author[1]{\fnm{B.} \sur{Mauri}}
\author*[2]{\fnm{V.} \sur{Mokina}}\email{valentyna.mokina@oeaw.ac.at}
\author[1]{\fnm{C.} \sur{Moore}}
\author[4]{\fnm{P.} \sur{Murali}}
\author[6]{\fnm{M.} \sur{Olmi}}
\author[8]{\fnm{T.} \sur{Ortmann}}
\author[6,13]{\fnm{C.} \sur{Pagliarone}}
\author[6,14]{\fnm{L.} \sur{Pattavina}}
\author[1]{\fnm{F.} \sur{Petricca}}
\author[8]{\fnm{W.} \sur{Potzel}}
\author[5]{\fnm{P.} \sur{Povinec}}
\author[1]{\fnm{F.} \sur{Pr\"obst}}
\author[6]{\fnm{F.} \sur{Pucci}}
\author[2,3]{\fnm{F.} \sur{Reindl}}
\author[8]{\fnm{J.} \sur{Rothe}}
\author[1]{\fnm{K.} \sur{Sch\"affner}}
\author[2,3]{\fnm{J.} \sur{Schieck}}
\author[8]{\fnm{S.} \sur{Sch\"onert}}
\author[2,3]{\fnm{C.} \sur{Schwertner}}
\author[1]{\fnm{M.} \sur{Stahlberg}}
\author[1]{\fnm{L.} \sur{Stodolsky}}
\author[7]{\fnm{C.} \sur{Strandhagen}}
\author[8]{\fnm{R.} \sur{Strauss}}
\author[4]{\fnm{F.} \sur{Toschi}}
\author[7]{\fnm{I.} \sur{Usherov}}
\author[2,3]{\fnm{D.} \sur{Valdenaire}}
\author[1]{\fnm{M.} \sur{Zanirato}}
\author[1,2]{\fnm{V.} \sur{Zema}}

\affil[1]{\orgname{Max-Planck-Institut f\"ur Physik}, \orgaddress{\postcode{85748}, \city{Garching}, \state{Germany}}}
\affil[2]{\orgname{MBI - Marietta-Blau-Institut f\"ur Teilchenphysik der \"Osterreichischen Akademie der Wissenschaften}, \orgaddress{\postcode{1010}, \city{Wien}, \state{Austria}}}
\affil[3]{\orgdiv{Atominstitut}, \orgname{Technische Universit\"at Wien}, \orgaddress{\postcode{1020}, \city{Wien}, \state{Austria}}}
\affil[4]{\orgdiv{Kirchhoff-Institute for Physics}, \orgname{Heidelberg University}, \orgaddress{\postcode{69120}, \city{Heidelberg}, \state{Germany}}}
\affil[5]{\orgname{Comenius University}, \orgdiv{Faculty of Mathematics, Physics and Informatics}, \orgaddress{\postcode{84248}, \city{Bratislava}, \state{Slovakia}}}
\affil[6]{\orgname{INFN}, \orgdiv{Laboratori Nazionali del Gran Sasso}, \orgaddress{\postcode{67010}, \city{Assergi}, \state{Italy}}}
\affil[7]{\orgname{Eberhard-Karls-Universit\"at T\"ubingen}, \orgaddress{\postcode{72076}, \city{T\"ubingen}, \state{Germany}}}
\affil[8]{\orgdiv{Physik-Department, TUM School of Natural Sciences}, \orgname{Technische Universit\"at M\"unchen}, \orgaddress{\postcode{85747}, \city{Garching}, \state{Germany}}}
\affil[9]{\orgdiv{Department of Physics}, \orgname{University of Oxford}, \orgaddress{\city{Oxford}, \postcode{OX1 3RH}, \state{United Kingdom}}}
\affil[10]{Also at: \orgdiv{LIBPhys, Departamento de Fisica}, \orgname{Universidade de Coimbra}, \orgaddress{\postcode{P3004 516}, \city{11}, \state{Portugal}}}
\affil[11]{Also at: \orgname{Walther-Mei\ss ner-Institut f\"ur Tieftemperaturforschung}, \orgaddress{\postcode{85748}, \city{Garching}, \state{Germany}}}
\affil[12]{Also at: \orgdiv{Instituto de F\'{i}sica}, \orgname{Universidade de S\~ao Paulo}, \orgaddress{\city{S\~ao Paulo}, \postcode{05508-090}, \state{Brazil}}}
\affil[13]{Also at: \orgdiv{Dipartimento di Ingegneria Civile e Meccanica}, \orgname{Università degli Studi di Cassino e del Lazio Meridionale}, \orgaddress{\postcode{03043}, \city{14}, \state{Italy}}}
\affil[14]{Also at: \orgdiv{Dipartimento di Fisica}, \orgname{Universit\`a di Milano Bicocca}, \orgaddress{\city{Milano}, \postcode{20126}, \state{Italy}}}

\abstract{
We present \impcresst{}, a \geant{}-based Monte Carlo tool to simulate backgrounds from natural and cosmogenic radionuclides, and calibration signals in solid-state detectors and their response to it. It is tuned for a fast-evolving and heterogeneous detector environment  with a focus on physics at the keV range. This tool was originally developed and validated by the CRESST collaboration; however, its flexibility and configurability make it suitable for other experiments with similar requirements. Key features of \impcresst{} include the dynamic geometry implementation directly from CAD files, ROOT-based data persistency of the whole event topology and automatic metadata annotation for data provenance, and interfaces to various particle generators, particularly for radiogenic and cosmogenic radionuclides. It includes also a newly developed particle generator for radioactive bulk and surface contaminations which is completely independent of any user defined confinement volumes. The auxiliary tool \cresstds{} applies detector-specific energy and time resolution based on a user-provided data set of empirical parameterization. We discuss also how to manage an \impcresst{} based workflow in an HPC environment based on Apptainer and nextflow. }


\maketitle

\section{Introduction}\label{sec:introduction}

We introduce \impcresst{}, a simulation package specially designed for solid-state detectors used to search for rare events, such as dark matter (DM) interactions, coherent elastic neutrino-nucleus scattering (CE\textnu{}NS), or neutrinoless double beta decay (0\textnu{}2\textbeta{}). It can be easily adapted to a particular experiment. To our knowledge, it is the only\footnote{Similar \geant{}-based tools either focus not on solid-state detectors (e.g.\ BambooMC \cite{Chen:2021} of the PandaX collaboration) or are not public available (e.g.\ MaGe \cite{Boswell:2011} of the Gerda and Majorana collaborations, QShields \cite{Adams:2024} of the \textsc{Cuore} collaboration, \textsc{Sage} \cite{She:2021} of the \textsc{Cdex} collaboration, Super\textsc{Sim} \cite{Agnese:2017} of the Super\textsc{Cdms} collaboration, or the \textsc{Nucleus} simulation tool \cite{Abele:2025}).} open source, \geant{} \cite{Agostinelli:2003,Allison:2006,Allison:2016,code:geant} based simulation package that focuses on the simulation of radioactive and cosmogenic backgrounds and calibration signals in cryogenic solid-state detectors and their response to it in the energy range \qty{100}{\eV}--\qty{10}{\MeV}. Beyond this focus, it may also be usable for other types of detectors in general. 
Starting with version 8.0.0 \cite{code:impcresst}, \impcresst{} will be available at 
\url{https://github.com/cresst-experiment/impcresst} as open source. This article outlines the capabilities and design choices of this package.

As the signal rate in rare-event searches is expected to be significantly smaller than the background, arising from well-known physical processes, any unknown background may limit the discovery potential of any kind of rare-event searches. Therefore, to extract the signature of new physics beyond the standard model of particle physics from the observation of rare events, background sources need to be well understood, making reliable background management essential. This includes the modelling of the background and its sources via Monte Carlo-based particle simulations for three broad use cases: decomposing background data into the individual background sources to determine their activities, predicting the amount of background which is to be expected in the physics data, and determining the background rejection power of new detector designs to inform the development process. To provide a particle simulation tool suitable for the \cresst{} direct DM search \cite{Angloher:2025a}, S.~Scholl originally developed \impcresst{} in 2011 \cite{Scholl:2011}. Albeit the \impcresst{} package is developed and maintained by the \cresst{} collaboration \cite{Angloher:2025a}, recent versions can be easily adapted to other experiments: the \textsc{Nucleus} collaboration used it for its initial background study \cite{Strauss:2017} and it is continuously used by the \textsc{Cosinus} collaboration for their background studies \cite{Angloher:2021,Fuss:2022}.

Since 2014 the \impcresst{} development was resumed and its design focused on six aspects, which are critical for \cresst{}: (i) easy handling of a diverse inventory of detectors' geometries and materials; (ii) intrinsic documentation and traceability of the provenance of the simulated data to avoid confusion between different detector designs; (iii) detailed recording of the simulated data to enable investigations of rare background processes or novel detector designs; (iv) aiming for the best possible accuracy at keV energies and down to the \qty{100}{\eV} scale; in future developments ideally down to the \qty{10}{\eV} scale; (v) offering interfaces to dedicated particle generators for cosmogenic and radioactive background sources, especially for radioactive bulk and surface contaminations; (vi) smooth integration in \emph{high performance computing} (HPC) environments.

With its unique feature set, \impcresst{} is a crucial part of \cresst{}'s roadmap for a next-generation DM search \cite{Angloher:2025a} and it was used for several publications  \cite{Turkoglu:2018,Scholl:2011,Fuss:2022,Fuss:2017,Abdelhameed:2019b,Angloher:2023,Angloher:2025,Angloher:2022a,Angloher:2024,Burkhart:2022}, which proved its suitability under the demanding environment of one of the leading experiments searching for low mass-DM \cite{Angloher:2024c}.
\impcresst{} feature set is also motivated by this environment, especially \cresst{}'s experimental setup and the background conditions. 

Using cryogenic calorimeters made of mono-crystals located deep-underground at the Laboratori Nazionali del Gran Sasso (LNGS) in Italy,  the \cresst{} collaboration has been pursuing a wide portfolio of physics cases: searching for nuclear recoils caused by spin-independent \cite{Angloher:2022,Angloher:2023b,Angloher:2024b,Angloher:2016,Angloher:2014,Abdelhameed:2019,Angloher:2024c} and spin-dependent \cite{Angloher:2022b,Abdelhameed:2019a,Angloher:2024c,Abdelhameed:2020} DM-nucleus scattering, Dark Photon-atom interactions \cite{Angloher:2017}, and DM self-interactions \cite{Angloher:2024a}. The continuous development of detector modules \cite{Angloher:2023a} over the years has resulted in a large inventory of detectors, requiring a flexible handling of simulated geometries, both in terms of geometry and target material:
massive $\mathcal{O}(\qty{100}{\gram})$-detectors \cite{Angloher:2016,Angloher:2014,Abdelhameed:2020,Bravin:1999}, light $\mathcal{O}(\qty{10}{\gram})$-detectors \cite{Abdelhameed:2019,Abdelhameed:2019a,Angloher:2024b}, wafer-like $\mathcal{O}(\qty{1}{\gram})$-detectors \cite{Angloher:2024b,Angloher:2022,Angloher:2024c}; featuring target crystals like \alo{} \cite{Bravin:1999}, \cawo{} \cite{Angloher:2024b,Angloher:2016,Angloher:2014,Abdelhameed:2019}, diamond \cite{Angloher:2023b}, \lialo{} \cite{Angloher:2022b,Abdelhameed:2020}, \ce{Li_2MoO_4} \cite{Abdelhameed:2019a}, and silicon \cite{Angloher:2022,Angloher:2024c}. Detector modules which use scintillating target crystals, like \cawo{}, are also equipped with a dedicated smaller cryogenic detector to determine the light yield for particle identification and discrimination. 
\cresst{} now routinely reaches detection thresholds for nuclear recoils below the \qty{100}{\eV}-scale \cite{Abdelhameed:2019,Angloher:2023b,Angloher:2022,Angloher:2024c}. With a silicon-on-sapphire wafer, a detection threshold as low as \qty{6.7}{\eV} was achieved, which enables us to set leading limits for the spin-independent, elastic DM particle-nucleon interactions down to $m_\mathrm{DM}=\qty{74}{\MeV}/c^2$ \cite{Angloher:2024c}.
Consequently, \impcresst{} focuses on a flexible handling of diverse detector geometries and of detector simulations at low energies.

Currently, the dominant background source for \cresst{} is radioactive contaminations of the detector targets, causing subsequent electromagnetic interactions \cite{Angloher:2024,Angloher:2022a,Abdelhameed:2019b}. 
Neutrons \cite{Fuss:2022}, either created by atmospheric muons or of radiogenic origin, and cosmogenic radionuclides \cite{Kluck:2021} are subdominant background components. However, the capture of thermal neutrons is relevant as calibration standard for nuclear recoils \cite{Thulliez:2021,Abele:2023} and was also observed in \cresst{}'s \cawo{} \cite{Fuss:2022,Angloher:2023} and \alo{} \cite{Angloher:2025} target crystals. Therefore, \impcresst{}'s modelling of physic interaction focus on radioactive decays and nuclear de-excitation.

The structure of the article follows a typical \impcresst{} simulation as done within \cresst{}'s simulation workflow (\cref{sec:workflow}). Each section corresponds to one aspect of the workflow: within the selected setup (\cref{sec:geometry}) the interactions (\cref{sec:physics}) of the particles from a background or calibration source (\cref{sec:particleGenerators}) are simulated. Data are extracted from the simulation (\cref{sec:particleTracking}) and stored (\cref{sec:dataformat}), onto which the detector response is applied  (\cref{sec:detectorresponse}). \Cref{sec:environment} gives an example of how this workflow can be implemented in a HPC environment with robust reproducibility and traceability. Finally, in \cref{sec:conclusion} we conclude and give an outlook on future developments of \impcresst{}.

For each section we will give the technical details of the implementation after we motivate them by the relevant physics case, and illustrate the practical usage on two examples\footnote{In the released \impcresst{} package, the examples can be found under \texttt{./examples/contaminantRange234Th/} and \texttt{./examples/decayPaths234Th/}, respectively.}: a simulation of a \ce{^{234}Th} contamination in a copper component of the \cresst{}'s setup as an \emph{extrinsic} background source and the range of its decay radiation (see \cref{fig:vertexPos}); and a \ce{^{234}Th} contamination of a \cawo{} target crystal as an \emph{intrinsic} background source and its decomposition in individual nuclear decay paths  (see \cref{fig:decay234}). For the convenience of the reader, the \impcresst{} settings to reproduce the first example are given in appendix \ref{sec:appendix}. The examples and the description in this article are based on the most recent \impcresst{} version 8.0.0, using \geant{} version 10.6.3.

\section{Workflow}
\label{sec:workflow}
Once an \impcresst{} user has decided \emph{what} to simulate -- what kind of background or calibration source (i.e.\ what kind of \emph{primary particle}) acting in what kind of experimental setup (i.e.\ what kind of \emph{geometry}) -- they can perform the simulation with the two tools of the \impcresst{} package: 
\impcresst{} itself for the \geant{}-based simulation of microscopic particle interaction and the accompanying \cresstds{} for applying the effects of finite detector resolution. Once the simulation pipeline is validated at small scales, one scales up the amount of primary particles for a high statistic \emph{production run} within an HPC environment (\cref{sec:environment}). For \cresst{} this can be done via \emph{lazy parallelism}, i.e., starting several simulation processes in parallel without any process communication between them, as we do not assume any correlation between the primary particles. Hence, \geant{}'s multi-threading capability is not yet used by \impcresst{}.

For a selected setup geometry (\cref{sec:geometry}) and based on the implemented physics model (\cref{sec:physics}), \impcresst{} simulates the \emph{microscopic interactions} of the primary particle (\cref{sec:particleGenerators}) with the materials out of which the chosen geometry consists. If the interaction occurs in a part of the geometry that corresponds to a detector, a \emph{hit} is created that encapsulates several data, most importantly the deposited energy (\cref{sec:particleTracking}). The hits, hereafter called \emph{raw hits}, along with additional data objects like particle trajectories or meta-data records, are stored as \emph{raw data} in ROOT file format (\cref{sec:dataformat})~\cite{Brun:1997,code:root}. \impcresst{} itself is controlled via \geant{} \emph{macro files} containing \emph{macro commands}~\cite[p.275]{geant4:application},  both native to \geant{} and specifically developed for \impcresst{}.

The tool \impcresst{} itself does not consider effects of the \emph{detector response}, i.e., the creation of signals from the particle interactions as they would be measured by the calorimeter and light detector with their finite time and energy resolution. To consider these, the raw data need to be processed to reconstruct energy depositions that are comparable to those measured by a real detector, hereafter called \emph{detector hits}, from the raw hits simulated with \impcresst{}\footnote{We note that in other experiments this is usually called \emph{event} reconstruction.}. This reconstruction is crucial to enable a direct comparison between simulated and real data. As the raw data are stored in ROOT files, a user can process them in any way supported by ROOT. A standard processing that applies only energy and time resolution is provided by the auxiliary \cresstds{} tool (\cref{sec:detectorresponse}). Due to the large variety of rather unusual materials (e.g.\ \cawo{}, \ce{LiAlO_2}), a microscopic description of the response of \cresst{}'s detector based on first-principles is not possible with satisfactory accuracy. Hence, \cresstds{} relies on empirical parametrizations of the detector resolutions obtained from analysing (calibration) data.

As typically the analysis pipeline changes faster than the geometry set-up of the experiment, this two-step approach saves computing time in exchange for temporary storage space: the time-consuming microscopic simulation with \impcresst{} needs to be done only once as soon as the geometry of the detector configuration is fixed. The simulated raw data can then be stored and processed multiple times to adapt to new analysis pipelines that may yield different effective detector resolutions or thresholds. Once the analysis pipeline converged to a stable state, the raw data can be deleted and only the size reduced processed data need to be stored, for an example of the attainable data reduction see \cref{sec:detectorresponse:data}.

As long as one keeps the macro file that controlled the simulation and the seed of the random number generator (see \cref{sec:particleGenerators:rng}  for details), one can \emph{reproduce} the raw data based on the \emph{traceability} of the simulated data: \impcresst{} follows a strict semantic versioning and the version number together with the used macro file is automatically noted in the metadata record with which each simulated data set is tagged. Using the same \impcresst{} version guarantees to use the same version of the implemented detector geometry. Reproducibility and traceability can be further enhanced by using \impcresst{} within a workflow that incorporates containerization and scientific pipeline management. As an example, \cref{sec:environment} describes the workflow implemented for the HPC environment used by \cresst{}.

\section{Geometry Definition and Implementation}
\label{sec:geometry}
The \impcresst{} package provides a flexible and extensible system designed to implement the geometry of the diverse configurations of detector modules and their supporting setup developed within the \cresst{} collaboration~\cite{Abdelhameed:2019,Abdelhameed:2019a,Abdelhameed:2019b,Abdelhameed:2020,Angloher:2014,Angloher:2016,Angloher:2022,Angloher:2023,Angloher:2023a,Angloher:2024}. The geometry subsystem builds upon the modular structure of \geant{} and enables users to select either built-in geometry configurations (\cref{sec:implemented_geometry}) or read them from CAD files with high precision (\cref{sec:cad_integration}), configure them (\cref{sec:dynamic_config}), and verify complex geometry layouts via visualisation (\cref{sec:visualization}).

The same package can be adapted for use in other cryogenic experiments employing similar detector technologies. A proof-of-principle was the usage by the \textsc{Cosinus} collaboration \cite{Angloher:2021,Fuss:2022}, hence the \impcresst{} codebase also includes legacy geometries originally used by them\footnote{All classes under the namespace \texttt{impCresst::geometry:: cosinus} correspond to geometries for an outdated prototype used by the simulation team of the \textsc{Cosinus} collaboration for background simulations. These geometries have since been updated to match
the final geometry utilized by the experiment, and results based on their simulation will be reported by the \textsc{Cosinus} collaboration in future
publications.}.  

\subsection{Implemented Geometry Types}
\label{sec:implemented_geometry}
\begin{figure}
    \centering
    \includegraphics{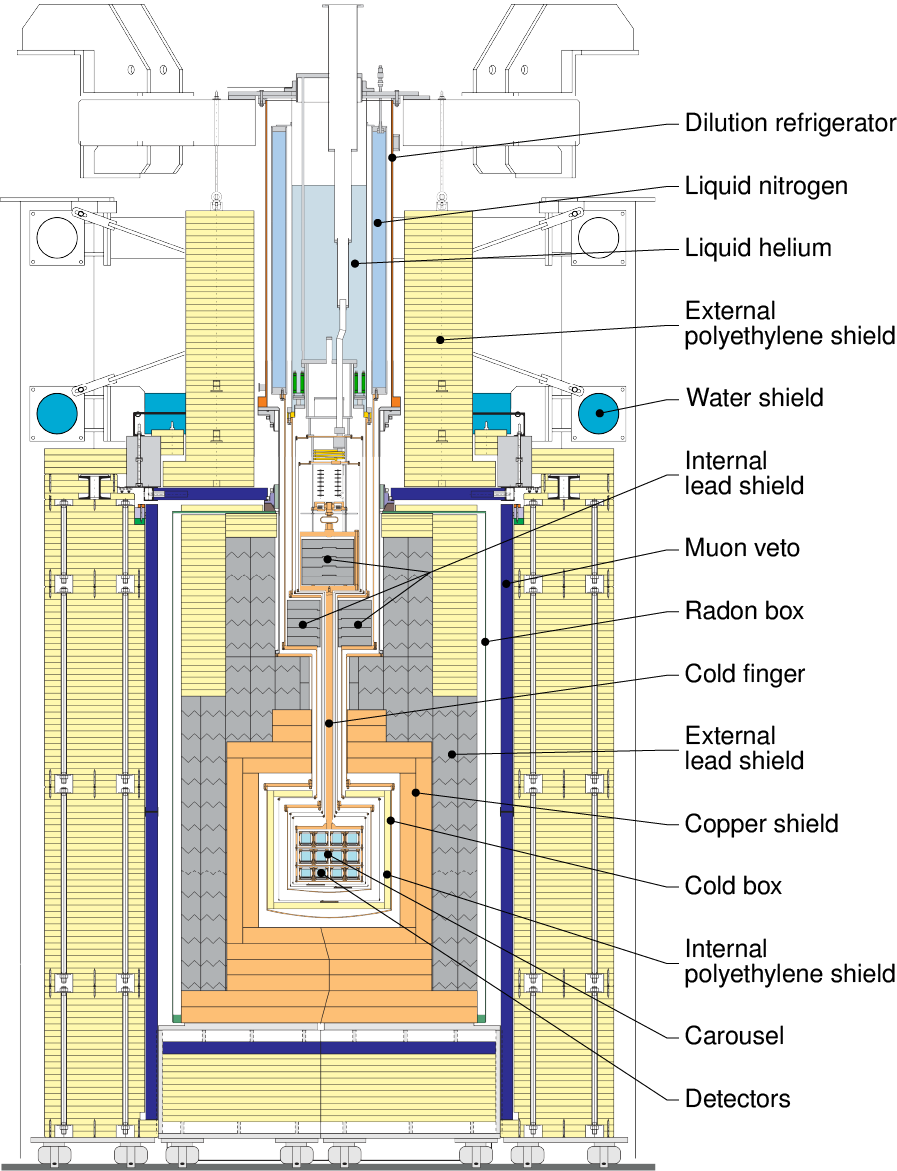}
    \caption{Technical drawing of the \cresst{} experimental setup. The cryostat holding the liquid helium and nitrogen can be seen at the top. Below the cryostat, the carousel hosts the detector modules. The shielding consists of polyethylene, lead, copper, an active muon veto, and the airtight radon box used to prevent radon contamination from air.}
    \label{fig:cresstSetup}
\end{figure}

The components of the \cresst{} experiment are arranged in a nested structure ranging from the inside to the outside, starting with individual cryogenic detector modules over an array of detector modules, the so-called \emph{carousel}, to the full experimental setup. As illustrated in \cref{fig:cresstSetup}, the setup consists of a muon veto and massive copper, lead, and polyethylene shielding against ambient background. The shielding encloses a cryostat that hosts the carousel where each detector module encapsulates one or more target crystals. 

\begin{figure*}
\centering
\includegraphics{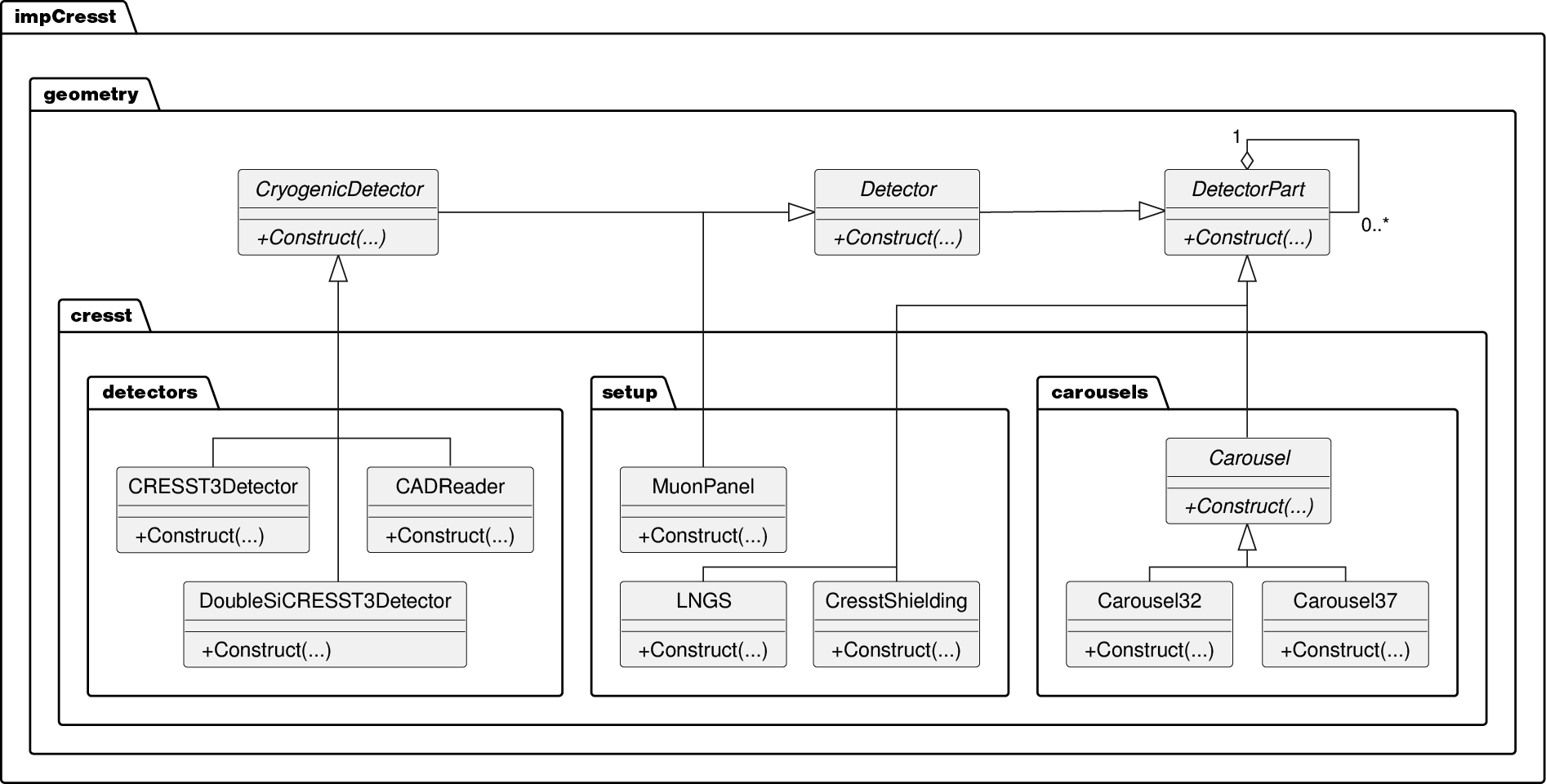}
\caption{\label{fig:geometry:detectorPartClass}Simplified UML class diagram of the abstract base class \texttt{DetectorPart} and derived classes to implement the hierarchy of components of the \cresst{} setup. For the sake of simplicity, the parameters and return type of the methods are omitted.}
\end{figure*}

\begin{figure*}
  \begin{center}
\begin{subcaptiongroup}
\begin{overpic}{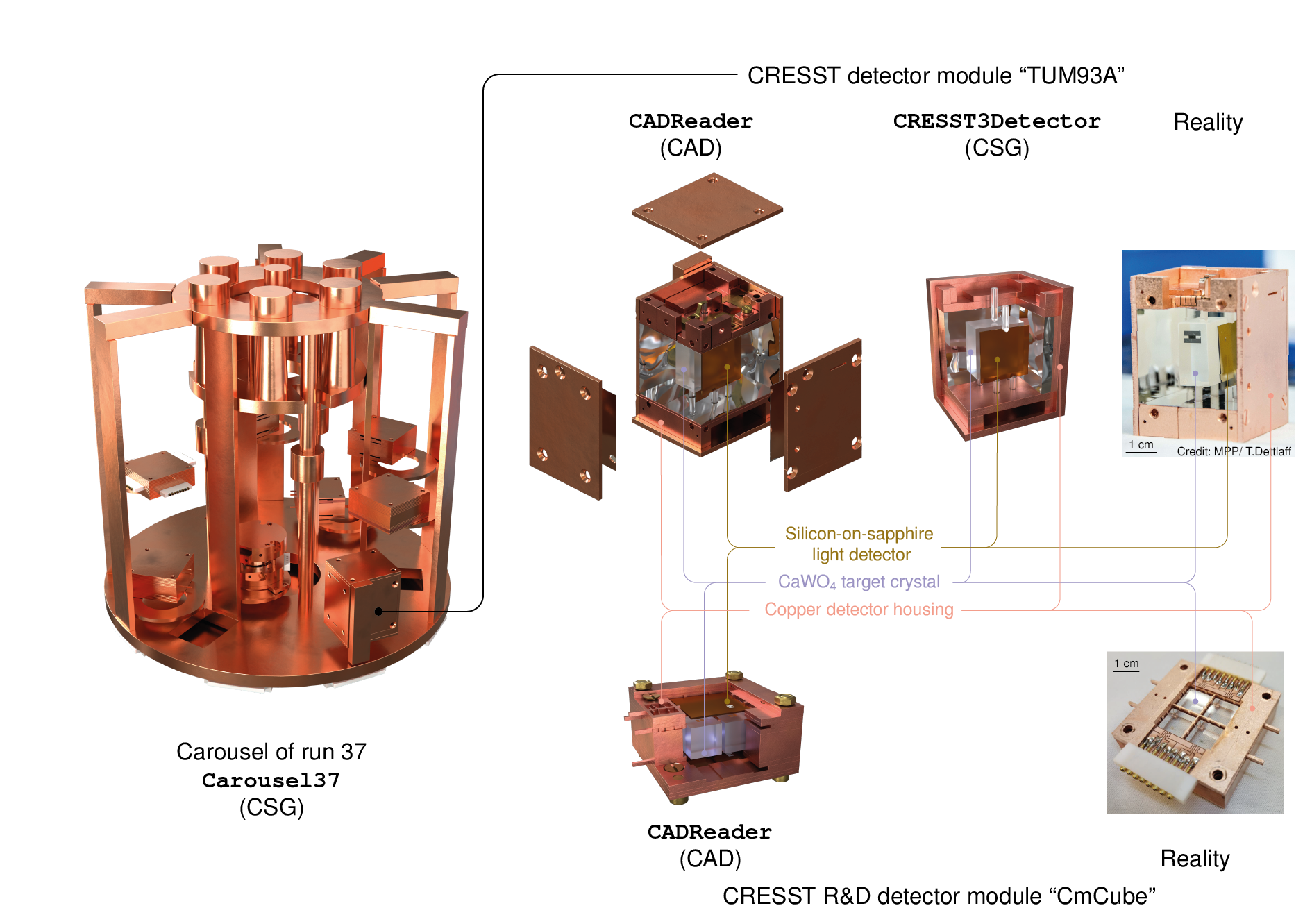}
	\phantomcaption
	\label{fig:vis:carousel37}
	\put(2.5,52){\captiontext*{}}
	\phantomcaption
	\label{fig:vis:tum93a:cad}
	\put(43,55){\captiontext*{}}
    \phantomcaption
	\label{fig:vis:cresst3Module}
	\put(67,55){\captiontext*{}}
	\phantomcaption
	\label{fig:vis:tum93a:pic}
	\put(85,55){\captiontext*{}}
	\phantomcaption
	\label{fig:vis:cmCube}
	\put(43,20){\captiontext*{}}
    \phantomcaption
	\label{fig:vis:cmCube:pic}
	\put(80,20){\captiontext*{}}
\end{overpic}
\end{subcaptiongroup}
  \end{center}
  \caption{Example visualizations with Blender of geometry components that are available in \impcresst{}:
  (\protect\subref{fig:vis:carousel37}) shows \texttt{Carousel37}, i.e., the detector array of the most recent data taking run 37. It itself is implemented as \emph{Constructive Solid Geometry} (CSG), but loads the detector modules from CAD files as objects of type \texttt{CADReader}, e.g., the ``TUM93A'' module (\protect\subref{fig:vis:tum93a:cad}, exploded view). Compared to a CSG implementation of the same module as instance of type \texttt{CRESST3Detector} (\protect\subref{fig:vis:cresst3Module}, cutaway view), the CAD implementation allows a higher fidelity with respect to the reality (\protect\subref{fig:vis:tum93a:pic}, photography of an open module). Loading geometries directly from CAD files also allows rapid simulations of R\&D projects like the potential future detector module ``CmCube'' (\protect\subref{fig:vis:cmCube}, cutaway view; \protect\subref{fig:vis:cmCube:pic}, photography of an open module) that may feature several target crystals kept in place within the detector housing only by their weight, see \cite{Angloher:2023a} for details. 
  \label{fig:vis}}
\end{figure*}

Within \impcresst{} these components are represented by subclasses that are derived from the abstract base class \texttt{DetectorPart}\footnote{In this text, we use simple, unqualified names when the class name is unique; if confusions are possible we mention also the (partial) namespace.} and that implement the virtual member function \texttt{Construct}, see the \emph{Unified Modeling Language} (UML, see e.g.\ \cite{uml}) class diagram shown in \cref{fig:geometry:detectorPartClass}. Each subclass implements the actual geometry of a component and its parts in the usual \geant{} way by using three types of \emph{volumes} \cite[p.97]{geant4:application}: solid, logical, and physical volumes. The geometrical shape and dimensions are represented via solid volumes of type \texttt{G4VSolid} which may be either a Constructive Solid Geometry (CSG) primitive or a tessellated solid (see \cref{sec:cad_integration}). These solid volumes are linked to materials via \texttt{G4LogicalVolume} objects, which are placed within the simulated world volume via \texttt{G4VPhysicalVolume} objects. Materials are defined within the \texttt{MaterialManager} class\footnote{We note, that we wrap the standard objects of type \texttt{G4Material}, that defines a material, and \texttt{G4VisAttribute}, that defines the visualisation of a logical volume, in our own structure \texttt{MaterialProperty} to ensure a consistent visualisation of a given material.}; materials used by \cresst{} are prefixed with \texttt{CRESST\_}.
Examples for the available types of \texttt{DetectorPart}s (form the outside to the inside) are:
\begin{itemize}
    \item \textbf{Setups:} objects of types \texttt{LNGS} and \texttt{CresstShielding} replicate the LNGS cavern and the full \cresst{} setup there, including movable shielding and cryostat.
    \item \textbf{Carousels:} objects of type \texttt{Carousel} implement detector arrays that represent multiple detectors and supporting structures. The geometry corresponds to specific data-taking runs and has to be matched to ensure direct comparability between simulation and data. \impcresst{} support the simulation of various runs from the legacy run 32 (\texttt{Carousel32}) to the most recent run 37 (\texttt{Carousel37}, see \cref{fig:vis:carousel37}).
    \item \textbf{Detector modules:} objects of type \texttt{CryogenicDetector} represent actual kinds of \cresst{} detectors that were used during data taking campaigns, such as \texttt{DoubleSiCRESST3Detector}~\cite{Angloher:2022}, \texttt{TUM40Detector} \cite{Angloher:2014} (used for simulations studies in \cite{Abdelhameed:2019b,Turkoglu:2018,Angloher:2024,Burkhart:2022,Angloher:2022a}), or \texttt{CRESST3Detector}~\cite{Abdelhameed:2019} (used in this work, see also \cref{fig:vis:cresst3Module}). An example for the latter type would be the detector module named ``TUM93A'' used in \texttt{Caraousel37}. In addition, dedicated calibration setups like \texttt{FeCalibration\_DoubleSi\_Left}~\cite{Angloher:2022} implement also the geometry of calibration sources, here \ce{^{55}Fe}, and are used for controlled studies of energy deposition, calibration, or background characterization.
\end{itemize}

\begin{figure}
\centering
\includegraphics{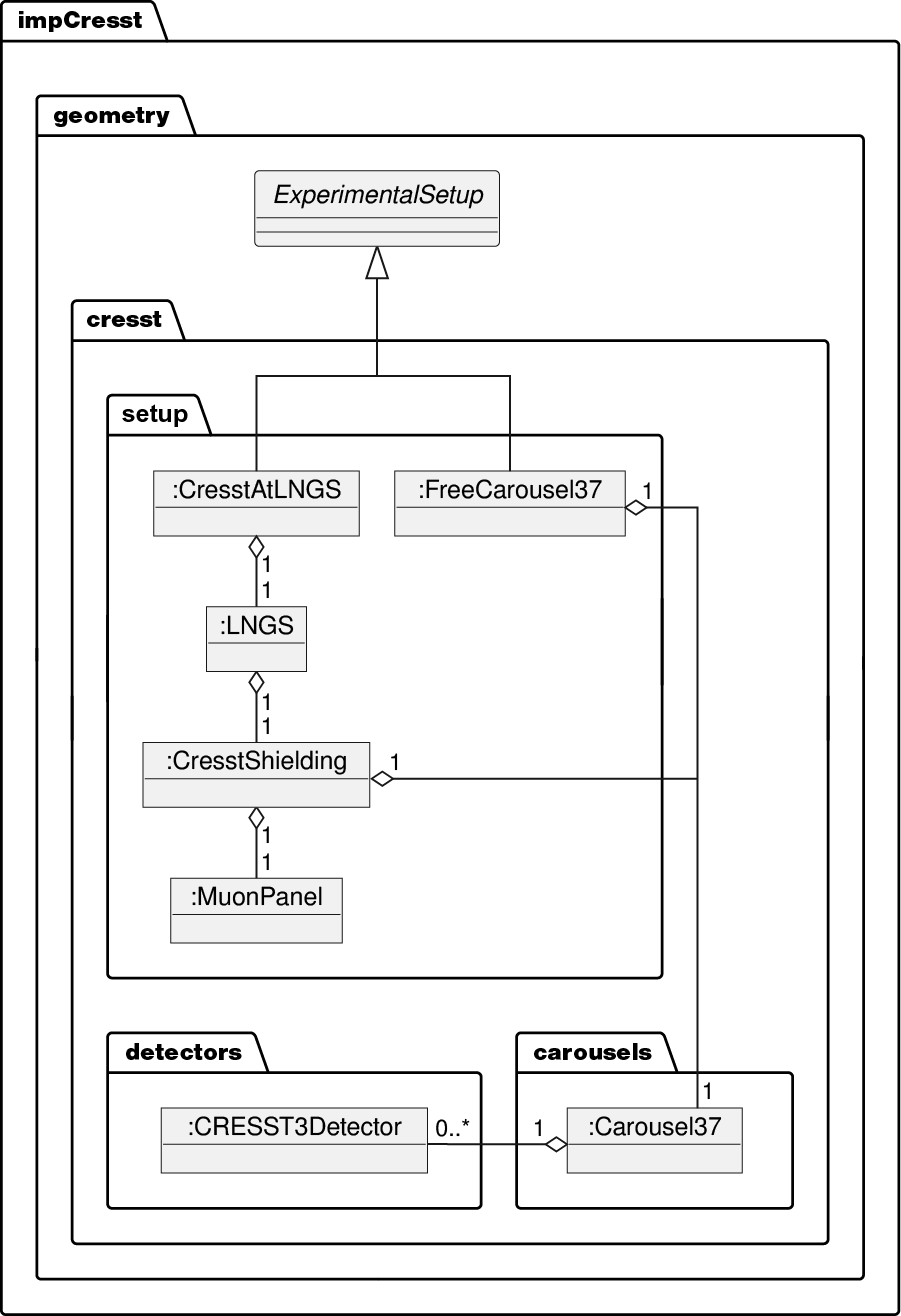}
\caption{\label{fig:geometry:detectorPartObject}Simplified UML object diagram illustrating the composition of two objects instantiated from subtypes of \texttt{ExperimentalSetup}: whereas \texttt{FreeCarousel37} encapsulates the composition of the carousel of data-taking run 37 with the related detector modules, \texttt{CresstAtLNGS} considered in addition the full experimental setup at LNGS. The detector modules have to be of a subtype of \texttt{CryogenicDetector}, e.g.,\ \texttt{CRESST3Module}.}
\end{figure}

Each object of type \texttt{DetectorPart} can be instantiated independently or combined to represent composite assemblies. Such assemblies are modelled by classes derived from the abstract base class \texttt{ExperimentalSetup}, see \cref{fig:geometry:detectorPartObject}. For example, the class \texttt{CresstAtLNGS} models the nested composition of \texttt{LNGS}, which contains \texttt{CresstShielding}, which contains one object that is instantiated from subtypes of  \texttt{Carousel}, which contains several objects that are instantiated from \cresst{}-specific subtypes of \texttt{CryogenicDetector} (see \cref{fig:geometry:detectorPartClass}). Complementary, \texttt{FreeCarousel37} omits all components outside the actual carousel. A use case for the former class would be the simulation of ambient backgrounds originating in the LNGS cavern, a use case for the latter class would be the simulation of short-ranged decay radiation from radioactive contaminants which will only be visible in the vicinity of the detector (like \ce{^{234}Th}, cf.\ \cref{fig:vertexPos}).

\begin{figure}
\centering
\includegraphics{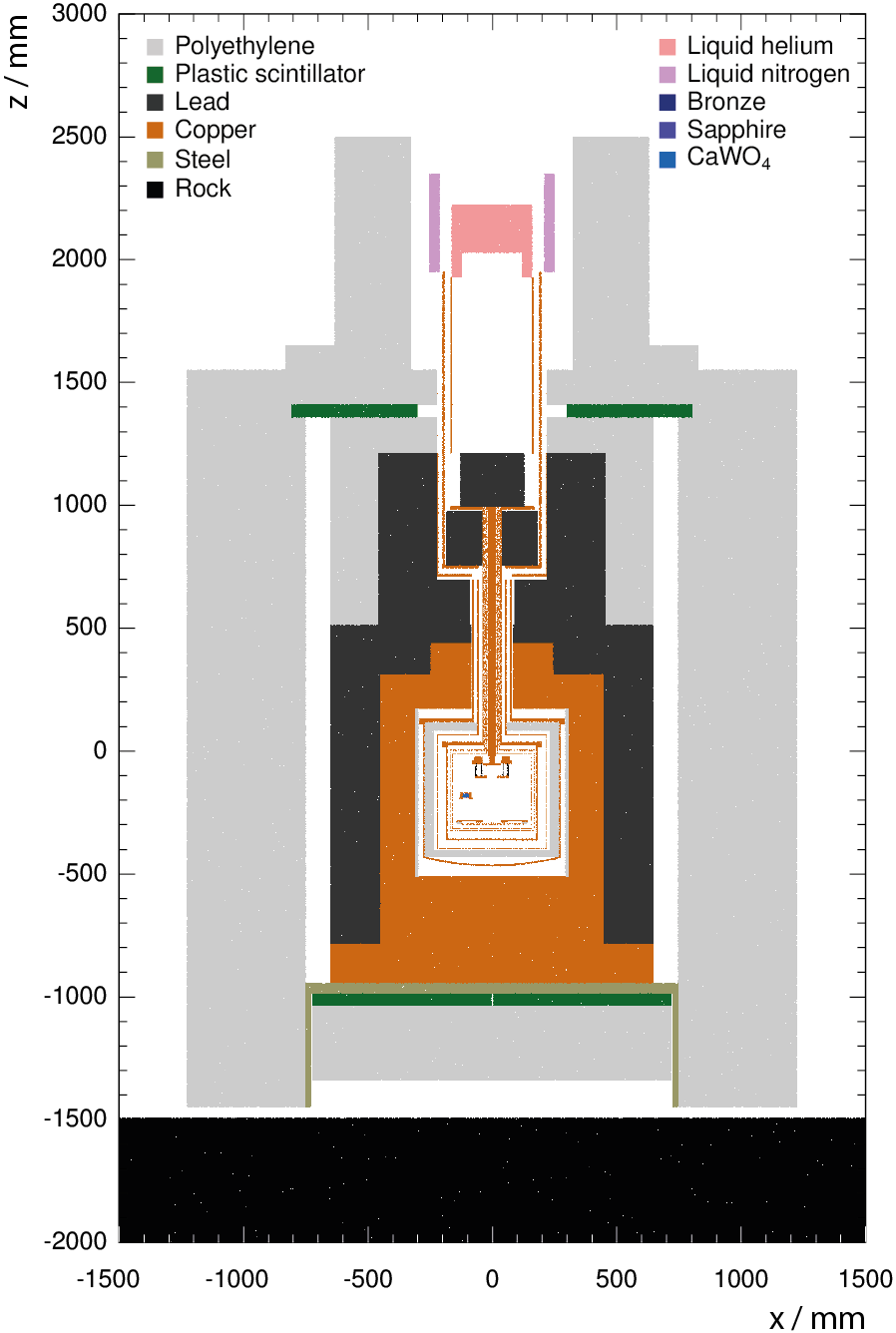}
\caption{Materials of the implemented \texttt{CresstAtLNGS} setup randomly sampled at \num{1e7} points in the $x$-$z$-plane. At $y=\qty{0}{\cm}$, the left and right panels of the muon veto have a tiny gap, needed to open the setup. Hence, these panels are not visible in this plot.}
\label{fig:material}
\end{figure}

Internal consistency checks are automatically performed at initialization to detect volume overlaps and boundary mismatches, ensuring that the simulated geometry conforms to the physical layout.
For further validation, users can generate 2D slices through the geometry via the macro command
\begin{lstlisting}
/geometry/createVolMap <y_pos> <x_half> <z_half> <n_points>
\end{lstlisting}
which produces files listing the material composition across the selected plane. These can be visualized as e.g., a scatter plot in ROOT (\cref{fig:material}). When compared to technical drawings (\cref{fig:cresstSetup}), it allows confirming material boundaries and detector interfaces, providing a quick verification tool complementary to graphical visualization (see \cref{sec:visualization}). Here, it shows that the implemented geometry of the upper part of the cryostat is simplified compared to the CAD drawing. This is justified, as the detectors are distant from the simplified parts and a massive internal lead shielding is placed in between.

\subsection{CAD Integration and Rapid Prototyping}
\label{sec:cad_integration}
A special subclass of \texttt{DetectorPart} is \texttt{CADReader} which provides a direct import of CAD geometries. Its main use case is to streamline the implementation of realistic detector models (cf. \cref{fig:vis:tum93a:cad}). Based on \texttt{CADMesh}~\cite{poole2011cadinterfacegeant4,code:cadmesh}, it reads CAD drawings in the Wavefront OBJ file format and models the described geometry as \texttt{G4TessellatedSolid} objects \cite[p.117]{geant4:application}. It can be easily selected via the corresponding \texttt{ExperimentalSetup} type \texttt{FromCADFile}. The macro command to achieve this is:
\begin{lstlisting}
/geometry/fromCADFile/ createCADReader <name> <obj file> <mapping file>
\end{lstlisting}
where the mapping file defines the correspondence between mesh names in the CAD file and material definitions within the simulation. It also specifies whether individual components are so-called sensitive detectors (see \cref{sec:particleTracking}). This way, a user can easily load any Wavefront OBJ compliant geometry in \impcresst{} and use it for simulation.

This functionality significantly accelerates geometry development and reduces potential translation errors. For example, the detector module geometries used in \cresst{} runs can be imported directly from engineering models, ensuring mechanical fidelity between simulated and real detectors. \Cref{fig:vis:tum93a:cad} shows a high fidelity model of the TUM93A detector module based on OBJ files, compared to a more simplified implementation based on CSG primitives shown in \cref{fig:vis:cresst3Module}. The CAD integration is particularly beneficial for rapid prototyping during detector design phases, where mechanical modifications can be immediately propagated into the simulation environment without the need of time-consuming manual coding of CSG primitives.
As example, \cref{fig:vis:cmCube} shows the visualization of a potential future ``CmCube'' detector module \cite{Angloher:2023a} which is made available in \impcresst{} via reading the CAD files\footnote{Within the \impcresst{} package, the user can find the CAD files under \texttt{./geometry/CmCubeDetector/}.}.

\subsection{Dynamic Configuration}
\label{sec:dynamic_config}
The geometry subsystem of \impcresst{} supports dynamic reconfiguration at runtime via macro commands \cite[p.275]{geant4:application}. Users can select and modify geometrical elements, such as changing materials, adjusting shielding positions, or substituting detector modules, without recompilation, e.g.,
    \begin{lstlisting}
/geometry/setSetup CresstAtLNGS
/geometry/CresstAtLNGS/ chooseCarousel Carousel37
/geometry/buildSetup
/geometry/CresstAtLNGS/Shift ShieldingRight 1.0 m
    \end{lstlisting}
selects the \texttt{ExperimentalSetup} subclass \texttt{CresstAtLNGS}. This subclass itself offers dedicated macro commands to choose a carousel, here the one  from run 37 as implemented in \texttt{Carousel37}, and to open the shielding of the setup to e.g., insert a calibration source (see also \cref{lst:exampleMacro}, lines~\ref{lst:macro:cresst}, \ref{lst:macro:carousel}, \ref{lst:macro:build}, \ref{lst:macro:shift}).
This flexibility enables efficient testing of alternative configurations, e.g., variations in shielding configuration or detector placement, during optimization studies.

\begin{figure}
\centering
\includegraphics{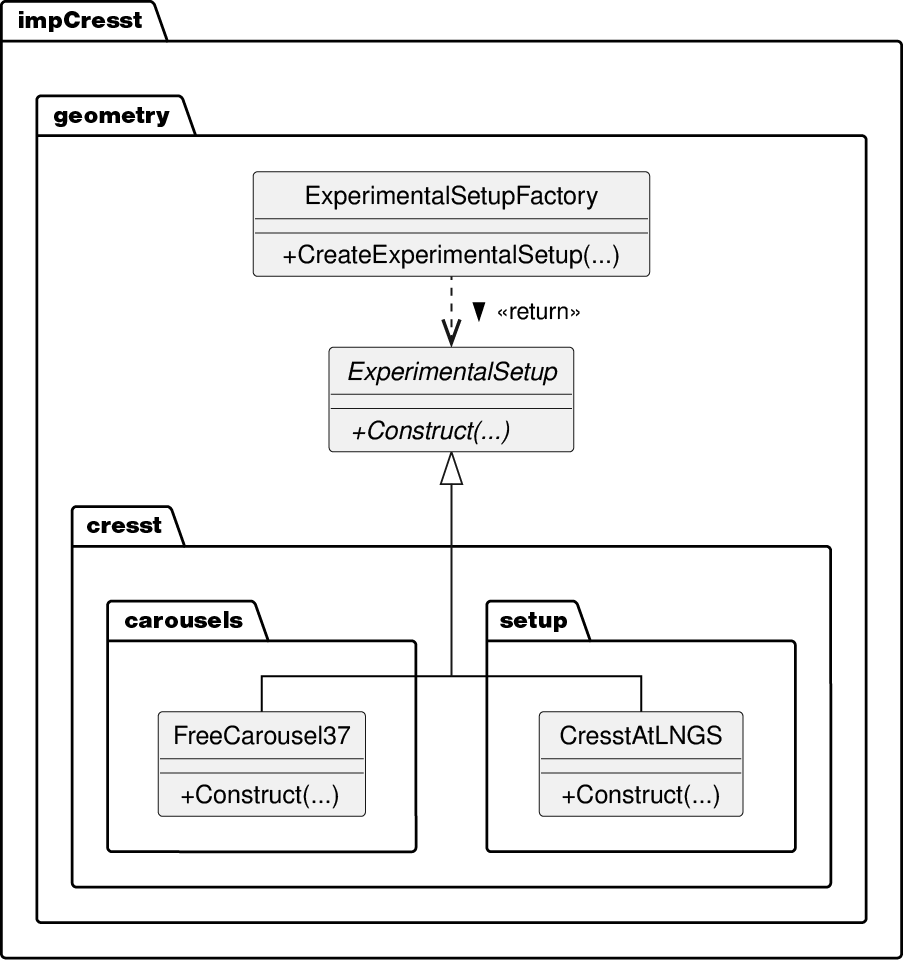}
\caption{\label{fig:geometry:factory} UML class diagram of the factory pattern used to return instances of types that implement the abstract base class  \texttt{ExperimentalSetup}, illustrated for the concrete types \texttt{CresstAtLNGS} and \texttt{FreeCarousel37}. For the sake of simplicity, the parameters and return type of the methods are omitted.}
\end{figure}

The selection of different types of \texttt{ExperimentalSetup}, like \texttt{CresstAtLNGS} or \texttt{FromCADFile}, are realized via the \texttt{ExperimentalSetupFactory} that implements the factory design pattern \cite{Gamma:2011}, see \cref{fig:geometry:factory}. Similarly, \texttt{CresstAtLNGS} calls an instance of \texttt{CarouselFactory} to create the actual carousel. \texttt{ExperimentalSetup} and \texttt{Carousel} calls an instance of \texttt{CryogenicDetectorFactory} to create the actual detector modules. This way, the user can easily specify the desired setup or detector component via string arguments to the respective macro commands and the factories instantiate the corresponding objects during runtime.

\subsection{Visualization Capabilities}
\label{sec:visualization}

Visualization is a crucial component of a simulation, serving both technical validation and presentation purposes. Because visualization data can become large, visualization is switched off by default in \impcresst{}. However, events of interest can be easily re-simulated, see \cref{sec:particleGenerators:rng}, with activated visualization. \impcresst{} makes use of \geant{}'s native visualization drivers \cite[p.285]{geant4:application} and external tools (need to be installed separately) such as \texttt{Jas3/WIRED4} \cite{code:jas3} and \texttt{Blender} \cite{code:blender}, providing a complete workflow for geometry verification, event inspection, and publication-quality rendering.

\impcresst{} supports multiple \geant{} visualization drivers\footnote{Within the \impcresst{} release, the macro files \texttt{./mac/visualization/visualizeGeometry.mac} and \texttt{./mac/visualization/vis.mac} illustrate how to select and configure the visualization drivers.}, including  \texttt{VRML} and \texttt{HepRepXML}, see \cite[p.285]{geant4:application} for a description. The latter is \impcresst{}'s default driver which allows the user to render detector geometries and event trajectories directly in \texttt{Jas3/WIRED4}; facilitating event-level inspection and geometry verification. 

For the production of publication-quality figures and outreach materials, we find the combination of \impcresst{} and \texttt{Blender} highly suitable.
Via the \texttt{VRML} visualisation driver, the loaded geometry is exported in the \texttt{VRML 2.0} \cite{vrml} format.
The resulting files can be inspected in external 3D viewers such as \texttt{ParaView} \cite{code:paraview} or directly rendered in \texttt{Blender}. This enables photorealistic visualization of detector components, see \cref{fig:vis} as examples.

In cryogenic rare-event searches such as \cresst{}, the accuracy of the geometry directly affects detector response predictions. Visualization tools in \impcresst{} have proven indispensable for confirming geometry integrity, inspecting imported CAD models, and communicating detector layouts to collaborators and the broader scientific community. The combination of event display, deterministic event replay, and advanced rendering ensures that simulation results are both technically verifiable and visually transparent.

\section{Physics Modelling}
\label{sec:physics}
As outlined in \cref{sec:introduction}, one set of requirements for \impcresst{} is to precisely model the surviving background, both its dominant and subdominant components, in the strongly shielded experimental set-up of \cresst{} at the keV energy scale. This background is dominated by electromagnetic interactions of radioactive contaminants \cite{Angloher:2024,Angloher:2022a,Abdelhameed:2019b}, whereas cosmic activation \cite{Kluck:2021} and neutrons (either from spontaneous fission, (\textalpha,n)-reactions, or induced by atmospheric muons) \cite{Fuss:2022} are a subdominant source of background for the current generation of \cresst{}; capturing of thermal neutrons is of interest for detector calibration \cite{Fuss:2022,Angloher:2023,Angloher:2025}. As \geant{} states the lowest applicability limit of its electromagnetic physics models \cite{geant4:applicabilityLimit} among the general purpose Monte Carlo codes\footnote{The other common general purpose Monte Carlo codes besides \geant{} are FLUKA \cite{Ballarini:2024,code:fluka} and MCNP \cite{Kulesza:2024,code:mcnp}.} we chose it for \cresst{}'s background simulations.
Following the recommendations of \geant{} for the use case of deep shielded experiments, we base the physics implementation of \impcresst{} on the ``Shielding'' physics list \cite[p.\,17]{geant4:physicslist} as provided with \geant{} version 10.6.3. Newer versions may offer different features, commands, and default settings.

In addition to the Shielding physics list, we prioritize high precision at the keV energy scale over simulation performance, which results in adjustments on the modelling of radioactivity (\cref{sec:physics:radioactivity}), electromagnetic interactions (\cref{sec:physics:em}), nuclear de-excitation (\cref{sec:physics:nuclearDeexcitation}), the production cut (\cref{sec:physics:productionCut}),  and optical interactions (\cref{sec:physics:optics}). We note, as \impcresst{} is open source and based on the open \geant{} framework, any user is free to further modify the physics list to their needs.

\subsection{Radioactivity}
\label{sec:physics:radioactivity}
By default, \geant{} implements radioactivity via the \texttt{G4RadioactiveDecayPhysics} physics constructor \cite[p.~403]{geant4:physics}, which reads \emph{Evaluated Nuclear Structure Data File}s (ENSDF) \cite{Tuli:1996}. In case the decay cause an excited daughter nuclei, nuclear de-excitation is implemented via the \texttt{G4PhotonEvaporation} class \cite[p.~349]{geant4:physics}.
Since \geant{} version 10.6, \texttt{G4RadioactiveDecayPhysics} relies on the \texttt{G4RadioactiveDecayBase} class that does not support biasing techniques any more \cite{geant4:release1060}. However, biasing may be relevant for future studies of e.g., the surviving ambient background from the rock around deep underground laboratories or the impact of cosmic activation as background in \cresst{}-like experiments. Hence, \impcresst{} uses the \texttt{G4Radioactivation} class that provides biasing in addition to the functionality of \texttt{G4RadioactiveDecayBase}. Furthermore, \geant{} treats tritium as stable by default. However, tritium is an important potential background source at the \qty{10}{\keV}-scale \cite{Angloher:2024}. Hence, we applied the modification described in Ref.~\cite{Kelsey:2020} to properly consider the tritium decay.

We note that we usually split the simulation of nuclear decay chains $X_i \rightarrow X_{i+1} \rightarrow X_{i+2} \rightarrow \dots$ in step-wise decays $X_i \rightarrow X_{i+1}, X_{i+1} \rightarrow X_{i+2}, X_{i+2} \rightarrow \dots$. For example, to stop the \ce{^{234}_{90}Th} decay once its daughter nuclides reach their ground state, we use the following \geant{} commands (cf.\ \cref{lst:exampleMacro}, line~\ref{lst:macro:grdm})
\begin{lstlisting}
/grdm/nucleusLimits 234 234 90 90
\end{lstlisting}
to configure \texttt{G4RadioactiveDecayBase}.
The splitting is possible because we can neglect the contribution to the real data of those radionuclides, e.g., \ce{^{214}Po} ($T_{1/2}=\qty{163}{\micro\s}$), that have half-lives shorter than the $\mathcal{O}(\qty{1}{ms})$-time resolution of \cresst{}'s detectors (see \cref{sec:detectorresponse:modelling}). Hence, there is no correlation between subsequent decays in the chain that has to be considered in the simulation. Therefore, we can simulate the energy deposition for each individual decay as normalized spectrum and scale it during post-processing to the measured absolute contamination activities or assuming a secular equilibrium of various degree. Hence, we do not have to re-run simulation for materials that differ only in their contamination activities. More details are given in \cite{Angloher:2024,Angloher:2022a,Abdelhameed:2019b}.

To precisely consider the atomic de-excitation following a radioactive decay, we usually run our simulation by enabling fluorescence, \emph{particle induced X-ray emission} (PIXE), and emission of Auger electrons via relevant \geant{} macro commands, see \cref{lst:exampleMacro}, lines~\ref{lst:macro:lowEnergyStart}-\ref{lst:macro:lowEnergyEnd}. To ensure that fluorescence photons and Auger electrons are always created as dedicated particle tracks, we ignore any production cut for atomic de-excitation processes. For PIXE cross-sections and fluorescence lines, we use the default data sets.

\subsection{Electromagnetic Interactions}
\label{sec:physics:em}
As electromagnetic interactions of radioactive contaminants are the dominant source of background for \cresst{}, the selected \emph{electromagnetic physics constructor} (EMPC) \cite[p.\,21]{geant4:physicslist} is of importance: it governs which physics models are used to implement the electromagnetic interactions. Our default EMPC is \texttt{G4EmStandardPhysics\_option4} which is regarded as the most accurate EMPC for low energy interactions\footnote{See \cite[p.\,211]{geant4:application}: \textquote{\texttt{G4EmStandardPhysics\_option4}, containing the most accurate models from the Standard and Low Energy Electromagnetic physics working groups.}}. Applied models relevant for radioactivity induced interactions are listed in \cref{tab:emop4:models}, for a full description see \cite[p.24]{geant4:physics}.  For the data-driven Livermore models, a lower applicability limit of \qty{250}{\eV} is recommended \cite{geant4:applicabilityLimit}. Albeit they stay operational at lower energies, their agreement with experimental data is not validated for our materials of interest. Currently, it is studied in a dedicated project how to validate \geant{} for energies below \qty{250}{\eV} in \cawo{} and \alo{}, and, if necessary, how to extend it \cite{Kluck:2023}.
\begin{table}[t]
\centering
\caption{Physics models applied in \texttt{G4EmStandardPhysics\_option4} according to \cite[p.24]{geant4:physics}. For each process, only the model that is applied at the lowest energies is listed. MCS stands for \emph{multiple Coulomb scattering}.}
\label{tab:emop4:models}
\begin{tabular*}{\linewidth}{@{\extracolsep{\fill}} l l}\toprule
Process & Model\\\midrule
e\textsuperscript{-}/e\textsuperscript{+} pair production & Penelope\\
Compton scattering & Monarsh University\\
Photo-electric effect & Livermore\\
Rayleigh scattering & Livermore\\
e\textsuperscript{-} MCS & Goudsmit-Sounderson\\
Ion MCS & Urban\\
Ion ionisation & ICRU73
\\ \bottomrule
\end{tabular*}
\end{table}

To easily select another EMPC, \impcresst{} provides a macro command, see \cref{lst:exampleMacro}, line~\ref{lst:macro:empc}.  Currently, \impcresst{} supports nearly all EMPCs provided by \geant{}\footnote{We support all EMPCs listed on \cite[p.\,21]{geant4:physicslist} except for \texttt{G4EmDNAPhysics} EMPC as it is out of scope for \impcresst{}.}. 

\subsection{Production Cut}
\label{sec:physics:productionCut}
Besides the chosen physics model, also the \emph{production cut} affects the simulation of particles: if the mean free path of a potential secondary particle is lower than the production cut, no secondary particle is created, but its kinetic energy is deposited locally. For a given material, this cut on the particle range can then be expressed as a cut on the particle energy \cite[p.\,204]{geant4:application}. A low production cut improves the simulation of \emph{leakage} at low energies i.e., the escape of low-energy particles that are produced near the surface of a detector which can reduce the energy deposited inside a detector. However, a low production cut increases the amount of simulated, individual secondary particles and, hence, increases the required computing time. Therefore, \impcresst{} makes use of production cuts per regions \cite[p.\,254]{geant4:application}: the farther away a geometric region is from the detector, the lower the probability that low energy particles from this region will reach the detectors. Hence, large production cuts are assigned to these far away regions (e.g.\ the region \texttt{Rock} containing the cavern of the LNGS laboratory), whereas lower production cuts are assigned to regions close to the detectors (e.g.\ the region \texttt{CryogenicDetectors} containing the detector modules themselves). Whereas the default production cut of \geant{} is \qty{0.7}{\mm} \cite[p.\,204]{geant4:application}, \impcresst{} use the values listed in \cref{tab:productionCuts}.

By default, the lowest energy to which the production cut can be converted is \qty{990}{\eV} \cite[p.\,204]{geant4:application}. In \impcresst{}, we lower this value down to \qty{250}{\eV}, the recommended applicability limits for the Livermore models \cite{geant4:applicabilityLimit}. For the detector volumes, we set a production cut of \qty{1}{\nm}: due to its limited dimensions, it is nearly guaranteed that it will always be translated to the lowest possible energy value of \qty{250}{\eV} regardless of the actual detector material. As nuclear recoils are the signature of the signal of interest for \cresst{} (see \cref{sec:introduction}), we ensure that they are always simulated by setting a production cut of 0 for protons, which acts on all recoiling ions.
\begin{table*}
\centering
\caption{\label{tab:productionCuts} Production cuts values applied in \impcresst{} for the various geometric regions. For the most prominent material per region, also the equivalent energy cuts for \textgamma{}-rays, electrons, and protons are given. The regions are ordered from most distant to closest to the detectors. For details, see text.}
\begin{tabular*}{\linewidth}{@{\extracolsep{\fill}} l S l S S S}
\toprule
\multirow{2}{*}{Region} & {\multirow{2}{*}{Production cut / \unit{mm}}} & \multirow{2}{*}{Material} & \multicolumn{3}{c}{{Energy cut / \unit{\keV}}}\\ \cmidrule{4-6}
       &   &   &  {\textgamma{}-rays} & {Electron} & {Proton} \\ \midrule
\texttt{Rock}               & 50 & GSRock & 49.12 & 27380 & 5000\\
\texttt{PEandPbShielding}   & 3 & Lead & 138.4 & 3963 & 300\\
\texttt{CuShielding}        & 1 & Copper & 24.74 & 1387 & 100\\
\texttt{MuonPanel}          & 0.5 & BC408 & 1.844 & 227.2 & 50\\
\texttt{Cavity}             & 0.1 & Copper & 7.320 & 249.8 & 10\\
\texttt{Carousel}           & 0.01 & Copper & 2.150 & 61.13 & 1\\
\texttt{CryogenicDetectors} & 1e-6 & \cawo{} & 0.250 & 0.250 & 0\\ \bottomrule
\end{tabular*}
\end{table*}

\subsection{Thermal Neutron Capturing}
\label{sec:physics:nuclearDeexcitation}
Although the kinetic energy of thermal neutrons ($E_\mathrm{n}=\qty{25}{\milli\eV}$) is below the detection threshold of most cryogenic detectors, at these energies the probability for neutron capture increases. In case of radiative capture, the emitted \textgamma{} rays are usually well above the detection threshold for \cresst{} and similar rare event searches. For example, the most prominent \textgamma{} line from \ce{_{20}Ca}(n,\textgamma{}) \cite{Reedy:2002} has an energy of $E_\mathrm{\gamma}=\qty{1.9}{\MeV}$. Beside this subdominant contribution to \cresst{}'s background budget \cite{Fuss:2022}, radiative neutron capture ${}^AX(\mathrm{n},\gamma)^{A+1}X$ can serve as calibration standard for nuclear recoils: e.g.,\ if only one \textgamma{} ray is emitted during the nuclear de-excitation, it will lead to a mono-energetic nuclear recoil of precisely known energy \cite{Thulliez:2021,Fuss:2022}. So far, \cresst{} observed this reaction in both \cawo{} \cite{Angloher:2023} and \alo{} \cite{Angloher:2025}.

Hence, the moderation of the neutrons down to thermal energies has to be accurately modelled. In case of \cresst{}, neutrons are mostly moderated by scattering with hydrogen nuclei inside the polyethylene shield inside and outside the cryostat. As advised in Ref.~\cite[p. 220]{geant4:application} we limit the \texttt{G4NeutronHPElastic} process \cite[p.~393]{geant4:physics}, which models the elastic scattering of neutrons with unbound target atoms, to neutron energies above \qty{4}{\eV}. For lower energies, we use the \texttt{G4NeutronHPThermalScattering} process \cite[p.~394]{geant4:physics}, which models the elastic neutron scattering with bound target atoms by considering dedicated cross-section data for thermal neutron scattering. In our case, we use the thermal cross-section data for hydrogen bound in polyethylene as provided by the \geant{} material \texttt{TS\_H\_of\_Polyethylene}. Both neutron scattering processes are data driven and use cross-section data from the \texttt{G4NDL4.6} compilation \cite[p.~393]{geant4:physics}, which relies mainly on the JEFF-3.3 \cite{Plompen:2020} and ENDF/B-VII.1 \cite{Chadwick:2011} databases.

For some gram-scale polyethylene parts inside the cryostat, the material temperature is on the $\mathcal{O}(\qty{10}{\milli\kelvin})$-scale and far below the lowest temperature for which \geant{} provides tabulated thermal cross-section data. In the case of \emph{inelastic} neutron scattering, \texttt{G4NeutronHPThermalScattering}\footnote{Actually, \texttt{G4NeutronHPThermalScattering} is an alias for \texttt{G4ParticleHPThermalScattering}.} of \geant{} version 10.6.3 attempts a linear extrapolation of the temperature dependent data to the mK-scale, which occasionally results in unphysical negative energies for the scattered neutron. 
As a workaround, we set the energy to 0 in this case. Because of the negligible mass of the affected parts, especially compared to the ton-scale polyethylene shield at room temperature, we expect no significant impact on the accuracy of the simulation.

By default, the precise description of nuclear recoils caused by radiative neutron captures is not possible as standard \geant{} is lacking the required accurate probabilities for single-$\gamma$ and multi-$\gamma$ emissions \cite{Thulliez:2021}, which constrained our previous studies in Refs.~\cite{Fuss:2022, Angloher:2023}. To remedy this, \impcresst{} in version 8.0.0 is using the \texttt{fifrelin4geant4} library \cite{f4g4}: it allows using the Fifradina dataset \cite{f4g4data} for the final state generation (i.e. emission of de-excitation \textgamma{} and recoiling daughter nuclei) for neutron capture on \ce{^{27}Al}, \ce{^{182}W}, \ce{^{183}W}, \ce{^{184}W}, and \ce{^{186}W}. This data set provides accurate emission probabilities taking into account energy and momentum conservation \cite{Thulliez:2021} based on FIFRELIN \cite{Litaize:2015} calculations. It also considers the impact of the slowing down of the recoiling daughter nuclei during the de-excitation process \cite{SoumSidikov:2023} via Iradina calculations \cite{Borschel:2011}.

\subsection{Optical Interactions}
\label{sec:physics:optics}
Typically, the scintillation light emitted by suitable target crystals and by the muon veto’s plastic scintillator is calculated during post-processing starting from the corresponding deposited energy using CresstDS (\cref{sec:detectorresponse}). However, the user may optionally use \geant{}'s \texttt{G4OpticalPhysics} class, which can be activated via (cf.\ \cref{lst:exampleMacro}, line~\ref{lst:macro:optics})
\begin{lstlisting}
/physics/activateOpticalPhysics true
\end{lstlisting}
to actually simulate the light emission and propagation. If activated, optical photons can be produced via scintillation or the Cherenkov process, and may undergo Rayleigh scattering, bulk absorption, and reflection at boundaries \cite[p. 207]{geant4:physics}. 
For reflection, the optical surfaces of the aluminium, bronze, and copper parts close to the target crystal are modelled as polished dielectric-metal boundaries. For \cawo{} target crystals, a grounded dielectric-dielectric interface is used with a roughness characterised by a slope distribution with a standard deviation of $\sigma_\alpha=\qty{6.7}{\degree}$. It is a  measured value after a surface treatment typical for \cresst{}-II target crystals \cite[table~2.9]{Sivers:2014}. As the plastic scintillator panels of the muon veto are wrapped in crumpled aluminium foil, a \texttt{groundair} finish together with a dielectric-dielectric interface is used. For the latter case, we use the \texttt{UNIFIED} model \cite{Levin1996} for the boundary processes, in all other cases the \texttt{GLISUR} model. 

In case of scintillation, we do not consider the quenching of the scintillation yield by the particle species (\textgamma, e\textsuperscript{-}, \textalpha, recoiling Ca, O, W nuclei) within the microphysics of the simulation as we could not reproduce the observed behaviour \cite{Fuss:2017}. Instead, we consider this via an empirical parameterization during the post-processing with \cresstds{} \cite{Fuss:2017}, see \cref{sec:detectorresponse}.

\section{Primary Particle Generation}
\label{sec:particleGenerators}
\impcresst{} allows the user to choose several ways to generate the primary particle\footnote{Within the released \impcresst{} package, examples of how to configure the primary particle generators via macro files can be found under \texttt{./mac/}.} (see e.g.\ \cref{lst:exampleMacro}, line~\ref{lst:macro:cosoSelect}), which can be selected from \geant{}'s default particle species \cite[p.~251]{geant4:application}. Besides \geant{}'s \emph{General Particle Source} (GPS) \cite[p.~22]{geant4:application}, we also interface particle generators that are more suitable for the simulation of cosmogenic background sources: CRY (\cref{sec:particleGenerators:cry}) and MUSUN (\cref{sec:particleGenerators:musun}). Within \impcresst{}, we also provide a particle generator especially developed for radioactive contaminations of surfaces and bulks: the \emph{\texttt{ContaminantSource}} (\cref{sec:particleGenerators:coso}). To ensure that subsequent runs of \impcresst{} are statistically independent, the particle generators and the subsequent particle tracking (\cref{sec:particleTracking}) are randomized using a pseudo \emph{random number generator} (RNG) that is initialized with distinct seeds at start up (\cref{sec:particleGenerators:rng}).

\subsection{Random Number Generator}
\label{sec:particleGenerators:rng}
As RNG we use RANLUX \cite{Luescher:1994,James2020,code:ranlux}\footnote{As provided in \geant{} via the  \texttt{CLHEP::Ranlux64Engine} implementation \cite[p.57]{geant4:application}.}, a high quality RNG suitable for MC simulations \cite{James2020}. We set its luxury level to the highest value of 4 and initialize it with the start time in nanoseconds\footnote{The time is obtained via the \texttt{std::chrono:: high\_resolution\_clock} available since C++11. Albeit its precision is machine dependent, precisions of at least \qty{1}{\ns} are obtained on our HPC nodes.} as seed. We safely assume that the number of simulation jobs started within the same nanosecond is negligible because even for massive parallel job submissions in HPC environments, the job scheduler needs a few tens CPUs cycles to start subsequent jobs. Hence, each run of \impcresst{} will have its unique seed, resulting in independent sequences of random numbers \cite{James:1994}. We decided against using (pseudo)random seeds\footnote{For example obtained via \texttt{/dev/urandom} on Unix-like systems or via hashing the time stamp.} as this may lead to the ``birthday problem'' \cite{cook:2016}, i.e., yielding the same seed value by chance.

\impcresst{} allows re-running individual events by storing the RNG state at the beginning of each event\footnote{If activated by the user via the \geant{} macro command \texttt{/run/storeRndmStatToEvent 1}, the RNG state is made available for each event. In this case, \impcresst{} obtains the state by calling \texttt{G4Event::GetRandomNumberStatus()} via \texttt{G4UserEventAction::BeginOfEventAction(const G4Event*)}.}. A use case would be for example to visualize (see \cref{sec:visualization}) one, particular interesting event selected out of a larger simulation run where visualization was switched-off to increase performance.  In case the user decides to re-run a particular event, they can recover the RNG state from the \texttt{Event} object stored in the raw data (see \cref{sec:dataformat}), export it to a simple text file\footnote{From the raw data, the RNG state can be exported for a selected event via \impcresst{}'s \texttt{Event::StoreRngStateToFile( const TString\& fileName)} function.}, and use this text file to re-seed the RNG at the beginning of the re-run (cf.\ \cref{lst:exampleMacro}, line~\ref{lst:macro:reseed}).

\subsection{CRY Interface}
\label{sec:particleGenerators:cry}
Particles produced in the atmosphere by primary cosmic rays \cite{Gaisser:2016} are an important background source for experiments at surface level or at shallow underground laboratories \cite{Heusser1995,Formaggio:2004}. For this use case, \impcresst{} can be linked against the \emph{Cosmic-ray Shower Library} (CRY) in version 1.7 \cite{Hagmann:2012,code:cry}. This allows the generation of atmospheric n, p, \textgamma{}, e\textsuperscript{-}, \textmu{}, and \textpi{} for three different altitudes and considering Earth's magnetic field and the solar activity, which affects the low energy-part of the particle spectra.

\subsection{MUSUN Interface}
\label{sec:particleGenerators:musun}
For experiments located at deep-underground laboratories, only atmospheric muons can penetrate the rock overburden while having significant interaction cross-sections \cite{Heusser1995,Formaggio:2004}. As the muon spectrum is affected by the energy loss caused by the passage through the rock, the actual overburden topography has to be considered. This is especially true, if the underground laboratory is placed under a mountain range with non-trivial topography, as it is the case for \cresst{} at LNGS below the Gran Sasso mountain range. For this use case, we developed an interface that allows \impcresst{} to read the output file produced by MUSUN \cite{Kudryavtsev:2009}. MUSUN is a particle generator specialized to calculate the muon flux at deep underground sites and provides the vertex data (position, direction, charge, and energy) of the muons after passing a rock overburden specified by a given topographic map. As MUSUN is an external program, it is the responsibility of the user to run it and to take care that the amount of calculated vertices matches the number of events to be simulated with \impcresst{}. If too few vertices are provided, \impcresst{} will wrap around and start again with the first vertex after issuing a warning.

\subsection{ContaminantSource}
\label{sec:particleGenerators:coso}
Finally, we developed a particle generator specialized in simulating the radioactive contamination of components of \cresst{}'s experimental setup, either their bulk volume or their surface area, with radionuclides: the \emph{\texttt{ContaminantSource}}. It is based on a three-step approach: (i) the user selects the components that should be “contaminated”, then \impcresst{} (ii) assigns a statistical weight to each of these components and (iii) samples the starting position of the primary particle vertices confined to these components.

In the first step, we provide macro commands to select components or their parts (implemented as objects of type \texttt{G4VSolid}, see \cref{sec:implemented_geometry}) based on their name, or the associated material, or the enclosing parent component. Technically, this is done by looking up the related object in \geant{}'s \texttt{G4PhysicalVolumeStore}. As an example (cf.\ \cref{lst:exampleMacro}, lines~\ref{lst:macro:cosoSelect}, \ref{lst:macro:cosoConfine}, \ref{lst:macro:cosoInit}), 
\begin{lstlisting}
/source/type contaminantSource
/contaminantSource/confine ToMaterialInVolume CRESST_Cu Carousel37
/contaminantSource/init
\end{lstlisting}
are the macro commands to select all copper components placed inside \texttt{Carousel37}.

In the second step, a statistical weight is assigned to each selected component: depending on the mode of \texttt{ContaminantSource}, bulk or surface contamination, this is either the component's bulk volume\footnote{Obtained via \texttt{G4VSolid::GetCubicVolume()}.} or its surface area\footnote{Obtained via \texttt{G4VSolid::GetSurfaceArea()}.} normalized to the total bulk volume or total surface area, respectively, of all selected volumes. As consistency checks during geometry construction prevents overlaps of detector parts (see \cref{sec:implemented_geometry}), relative bulk volumes and relative surface areas are suitable statistical weights.

In the final step, we sample a starting position from the selected components in two steps: as both volumes and surfaces are additive, the probability distribution for the starting position is factorized, and one can first sample one component and afterwards a random point within the bulk or on the surface of this component.
The sampling of the component is implemented as \emph{inverse transform sampling} \cite{Devroye1986} based on the \emph{cumulative distribution function} (CDF) of the statistical weights.

In surface mode, \texttt{ContaminantSource} obtains a random point on the component's surface via \texttt{G4VSolid::GetPointOnSurface()}. As \geant{} does not offer a similar functionality to obtain a random point in the bulk of a \texttt{G4VSolid} object, we manually obtain one via \emph{rejection sampling} \cite{Devroye1986}: we circumscribe the selected component, which may have arbitrary shape, with a \emph{bounding box}\footnote{The dimensions of the bounding box are set to the maximum value obtained via \texttt{G4VSolid::BoundingLimits( G4ThreeVector min, G4ThreeVector max)}.}. Afterwards, we homogeneously sample points from the bounding box and keep only those that are actually inside the selected component\footnote{We use \texttt{G4Navigator::LocateGlobalPointAndSetup( G4ThreeVector position, G4ThreeVector direction)} to obtain the volume that encompasses the sampled \texttt{position}.}. We take into account the transformation between the local coordinates of the selected volume and \geant{}'s global coordinates of its \emph{world volume}, which we have to sample to obtain a suitable starting position\footnote{The selected volume may be part of a nested volume hierarchy rooted at the world volume. \texttt{ContaminantSource} is reverse traversing the hierarchy, starting at the selected volume aiming towards the world volume, by accessing the mother volume of a given volume via the \texttt{G4PhysicalVolumeStore}. At each traversing step, any geometrical transformation applied between the actual mother volume-daughter volume pair is obtained via \texttt{G4VPhysicalVolume::GetObjectTranslation()} and \texttt{G4VPhysicalVolume::GetObjectRotationValue()}. Concatenating the individual transformations yields the total transformation.}. 

In surface mode, the user may also consider the \emph{implantation depth} $x$ of the surface contaminants, i.e., the extent to which the starting position is randomly shifted below the surface. Currently, three two-parametric models are implemented, all accepting an offset $A$:
(i) an \emph{exponential distribution} $A-\exp(-x/\lambda)$, where $\lambda$ is its characteristic length;
(ii) a \emph{Gaussian distribution} $G(x;A,\sigma)$, where $\sigma$ is its standard deviation;
(iii) a \emph{uniform distribution} $A+x_\mathrm{max}\cdot U(x)$, where $x_\mathrm{max}$ is the maximum of its value range.

The user can set the particle species and kinetic data of the start vertex of the primary particle either by using the GPS or, in case of neutrons, via an interface to SOURCES4 \cite{Wilson:2005}. As an example for GPS usage, the following commands select \ce{^{234}Th} at rest as contaminant (cf.\ \cref{lst:exampleMacro}, lines~\ref{lst:macro:gpsFirst}-\ref{lst:macro:gpsLast}):
\begin{lstlisting}
/gps/particle ion
/gps/ion 90 234
/gps/ang/type iso
/gps/energy 0 MeV
\end{lstlisting}
This non-trivial decay has a significant effect on the hit reconstruction for \cresst{}, see \cref{sec:detectorresponse}.

For neutrons, \texttt{ContaminantSource} is able to sample the kinetic energy of neutrons from the output file of a neutron yield calculation done via SOURCES4. The initial neutron direction is assumed to be isotropic in this case.

\begin{figure*}[t]
\centering
\includegraphics{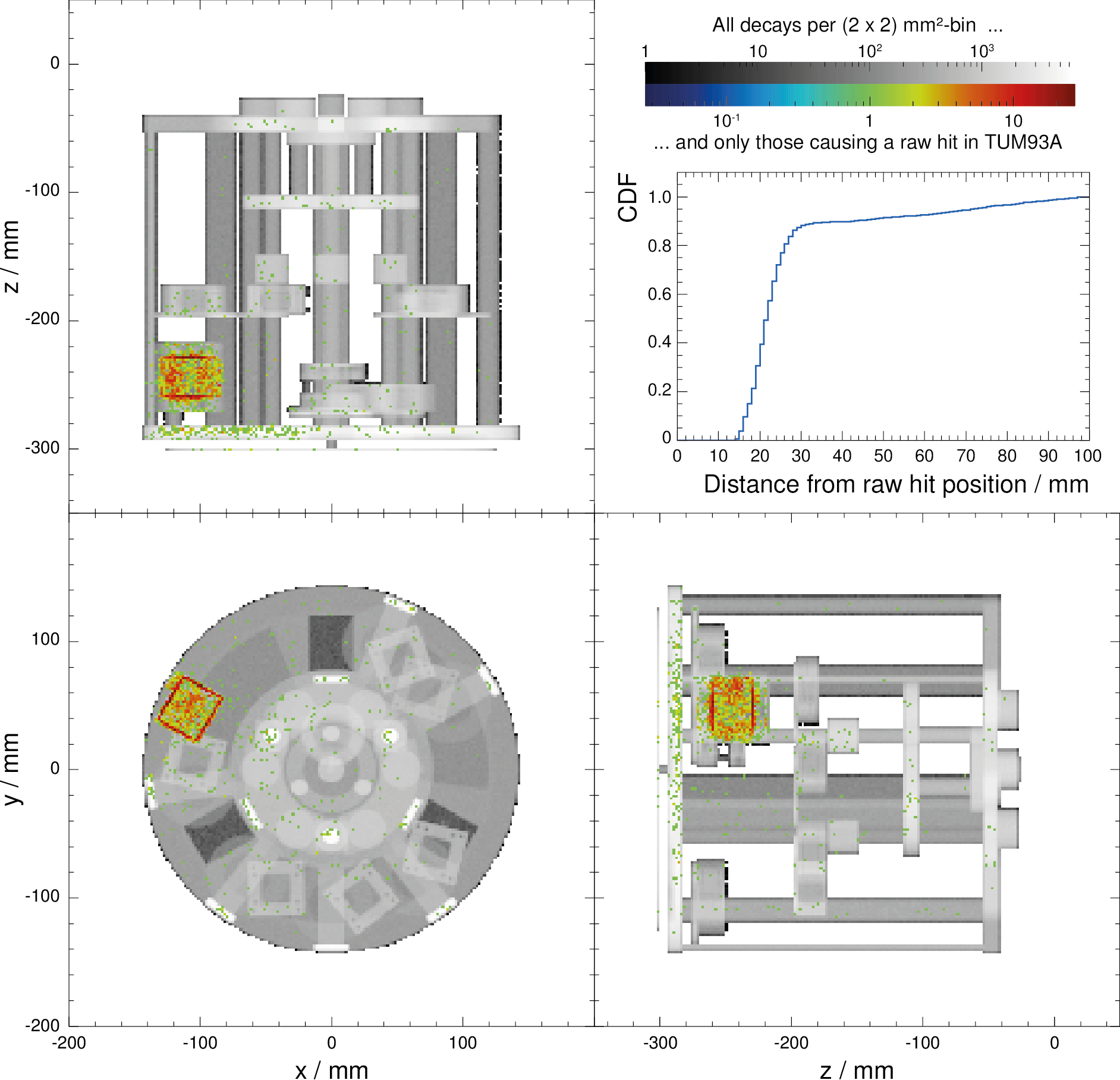}
\caption{\label{fig:vertexPos}Simulation of \num{1e7} decays of \ce{^{234}Th} placed in all copper parts of \cresst{}'s \texttt{Carousel37}. The three-way projection shows the vertices of the primary particles in logarithmic \emph{grey scale}. The logarithmic \emph{colour scale} shows the vertices of those \num{2351} primary particles that cause an energy deposition, i.e., a \emph{raw hit}, inside the \cawo{} target crystal of the TUM93A detector module. The \emph{Cumulative Distribution Function} (CDF) shows that most primary particles that cause an energy deposition originate less than \qty{3}{\cm} away from the hit position.}
\end{figure*}

Based on the example used in this section, \cref{fig:vertexPos} illustrates the result of using \texttt{ContaminantSource} in bulk mode: the \emph{grey-scale} images show the starting vertices of \ce{^{234}Th} constrained to the copper parts of the \texttt{Carousel37} geometry. Comparing \cref{fig:vertexPos} with \cref{fig:vis:carousel37} shows that the vertices are clearly confined to the copper parts, which confirms the correctness of the sampling process.

\section{Particle Tracking}
\label{sec:particleTracking}
With the generation of a primary particle, a \geant{} event (type \texttt{G4Event}) is created: this encapsulates all subsequent actions like particle interactions or secondary particle production, until all particles either have left the simulated \emph{world volume}, decay, or stop.
In the following, the current state of a particle trajectory is represented by a \emph{track} (type \texttt{G4Track}) object and its propagation happens in individual \emph{steps} (type \texttt{G4Step}).
To extract information from the particles along their track through the implemented geometry, \impcresst{} use objects of base type \texttt{G4VSensitiveDetector} and \texttt{G4VPrimitiveScorer}.

\emph{Sensitive detectors} \cite[p.173]{geant4:application}, derived from the class \texttt{G4VSensitiveDetector}, represent the abstract concept of detector-like behaviour, i.e., the sensitivity of an experimental component to energies deposited via interactions with the component's material. Consequently, it is only sensible to attach them to geometric volumes that are linked to a material, i.e., that are part of the \emph{mass geometry} \cite[p.73]{geant4:application} described in \cref{sec:geometry}. The energy depositions are represented as subtype of \texttt{G4VHits}. Within \impcresst{} we use this framework to model two types of detectors: (i) the plastic scintillator panels of \cresst{}'s muon veto as sensitive detectors of type \texttt{PlasticScintillator}, creating hits of type \texttt{PlasticScintillatorHit}; and (ii) the target crystals and light detectors inside the cryogenic detector modules as sensitive detectors of type \texttt{Bolometer}\footnote{We note that \cresst{}'s cryogenic detector may operate also in \emph{calorimetric} mode. However, from the point of view of the simulation there is no difference to the \emph{bolometric} mode, hence we do not represent these two modes in the software design.}, creating hits of type \texttt{BolometerHit}. 

To extract data unrelated to interactions with a material, e.g., the flux of particles passing through the surface of a geometric volume, we use \emph{scorers} \cite[p.173]{geant4:application} derived from class \texttt{G4VPrimitiveScorer}. They are placed in \geant{}'s \emph{parallel world} \cite[p.184]{geant4:application}: overlaying the mass geometry, it is an alternative hierarchy of volumes, which does not have to be linked to materials. In this case, a particle can propagate through volumes in the parallel world without interacting. This makes it ideal to measure a particle flux without affecting the particles during measurement.
By default, \geant{} offers a variety of possibilities for this so-called \emph{scoring}, e.g., defining parallel world volumes and associated scorers via macro commands \cite{Asai:2019}. However, it lacks a convenient way to attach a scorer in the parallel world to an existing component in the mass world without entering the position and dimensions of the volume by hand. \impcresst{} provides this feature: 
macro commands are available with which the user can select a component in the mass world and attach a primitive scorer to it that is able to extract particle current, flux, or energy while it passes through the surface of the component. For example, to score the energy of incident gamma rays through all surfaces of the volume \texttt{TUM93A\_CaWO4}, i.e., the \cawo{} target crystal of the detector module \texttt{TUM93A} of \texttt{Carousel37}, the following macro commands are sufficient (cf.\ \cref{lst:exampleMacro}, lines~\ref{lst:macro:scorerFirst} to \ref{lst:macro:parallelWorld}):
\begin{lstlisting}
/geometry/addScorer TUM93A_CaWO4 gamma energy 0 in
/geometry/buildParallelWorld
\end{lstlisting}
The technical implementation is straightforward: the mass world volume related to the selected component is retrieved via \texttt{G4PhysicalVolumeStore} and a volume of the same shape and dimensions is created in the parallel world at run time. Subsequently, a scorer implementing the abstract base class \texttt{\texttt{G4VPrimitiveScorer}} is attached. As the correct calculation of a flux depends on the actual shape of the volume, the scoring is currently limited to volumes of type \texttt{G4Box}, \texttt{G4Tubs}, \texttt{G4Sphere}.

By subclassing \geant{}'s \emph{user action classes} \cite[p.~268]{geant4:application}, mainly \texttt{G4UserEventAction} and \texttt{G4UserTrackingAction}, \impcresst{} extracts not only the hits registered by sensitive detectors and the values measured by scorers, but also the full \emph{event topology}: the connections between the primary particle, its secondary particles, and the hits caused by them. Storage of these data is described in \cref{sec:dataformat}.

\section{Persistency of Raw Data}
\label{sec:dataformat}
\impcresst{} version 8.0.0 uses ROOT \cite{Brun:1997} version 6.32.08 to stream the simulated data to files. Following \cref{sec:workflow}, the actual layout of the raw data (see \cref{sec:dataformt:layout}) has to be detailed, as it serves as input for later post-processing (\cref{sec:detectorresponse}). Furthermore, we aim to decouple the dependencies of data simulation with \geant{} and data processing with ROOT which also impacts the design of the related data containers (\cref{sec:datalayout:decoupling}). Ways to minimize the size of the produced files are discussed in \cref{sec:datalayout:size}. Finally, we annotate our simulated data with metadata for data provenance (\cref{sec:datalayout:metadata}).

\subsection{Data Layout}
\label{sec:dataformt:layout}
We aim to store the complete event topology. This includes especially the relations between particle tracks and their ``parent track'' and ``child tracks'', as well as the relation between a particle track and the hits it caused. For this reason, we use a monolithic event class \cite{code:root:CustomClasses} which acts as leaf for a ROOT \texttt{TTree} \cite{code:root:TTree} stored inside a ROOT file of type \texttt{TFile} \cite{code:root:TFile}. \impcresst{}'s data interface\footnote{These classes are encapsulated in the name space \texttt{impCresst::dataRecord::interfaces}.} interacts with \geant{}'s \emph{User Action Classes} \cite[p.263]{geant4:application}, sensitive detectors, and scorers (\cref{sec:particleTracking}) to extract the data and pass it to dedicated container classes which are aggregated by our ROOT event class\footnote{These classes are encapsulated in the name space \texttt{impCresst::dataRecord::containers}.}.

\begin{figure*}
\centering
\includegraphics{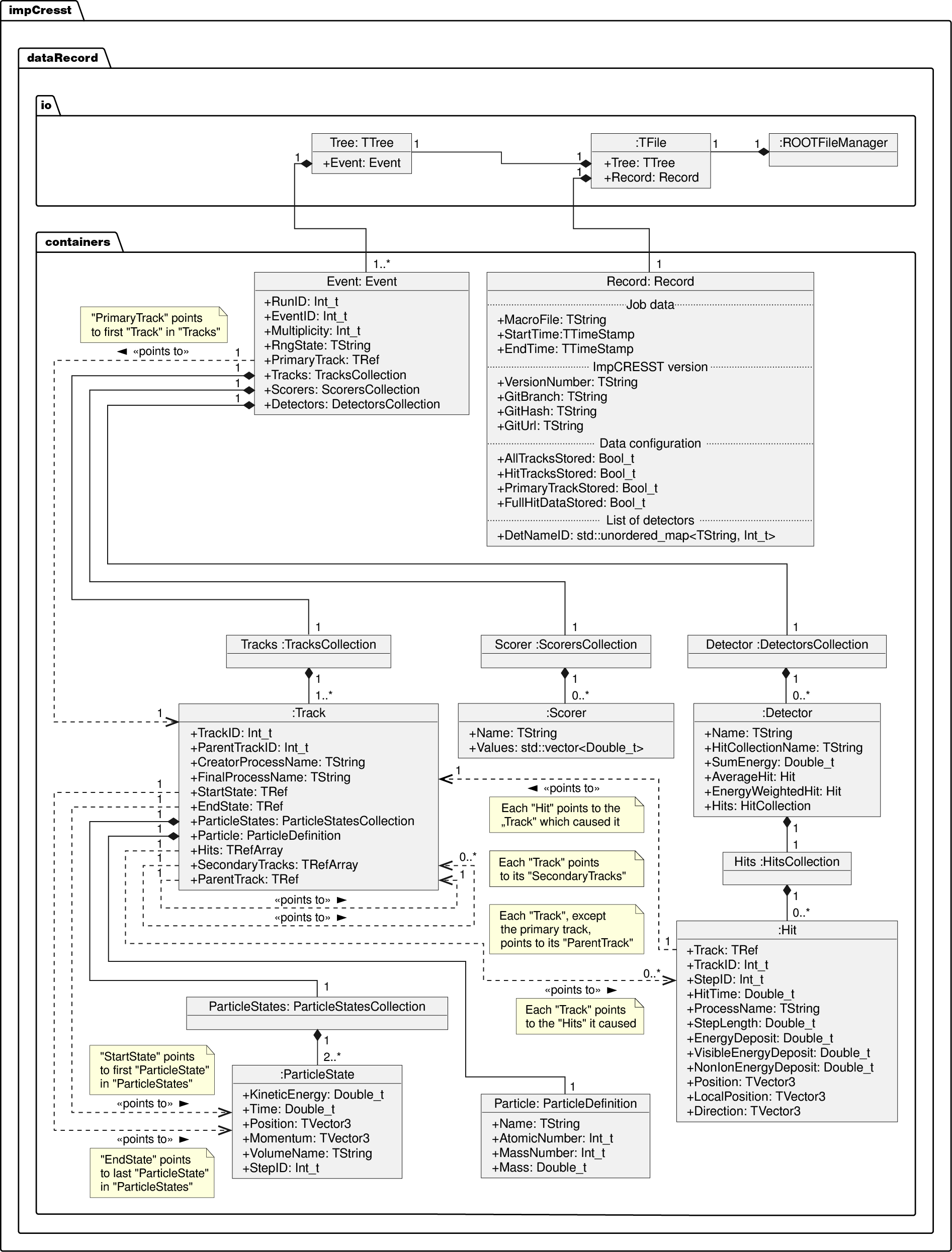}
\caption{\label{fig:rawFileFormat}Simplified UML object diagram of the data structure used by \impcresst{} to store the raw data of the simulation in ROOT files. For the sake of simplicity, not all data members of \texttt{Record} are shown. For details, see text.}
\end{figure*}

\Cref{fig:rawFileFormat} illustrates the resulting data layout as UML object diagram \cite{Fowler:2003} and indicates the most relevant of the stored data\footnote{For a complete list see the doxygen \cite{code:doxygen} API documentation that can be generated during installation of \impcresst{}.}. As one can see, we mirror most of the relevant \geant{} data classes with our own container classes: \texttt{G4VHit}/\texttt{Hit}; \texttt{G4Track}/\texttt{Track}, \texttt{G4Event}/\texttt{Event}, \texttt{G4Step}/\texttt{ParticleState}, \texttt{G4ParticleDefinition}/\texttt{ParticleDefintion}, and \texttt{G4VSensitiveDetector}/\texttt{Detector}, respectively.
By using ROOT's \texttt{TRef} \cite{code:root:TRef} and \texttt{TRefArray} \cite{code:root:TRefArray} types we realize lightweight, persistent references between different \texttt{Track} and \texttt{Hit} objects. For example, each hit points to the track that creates it and vice versa, or we provide with \texttt{PrimaryTrack} a shortcut to the \texttt{Track} object stored in the \texttt{Tracks} collection that belong to the primary particle.

Besides the raw \geant{} data, three derived entities are provided for each \texttt{Detector} class: \texttt{SumEnergy}, \texttt{AverageHit} and \texttt{EnergyWeightedHit} to give quick access to the total energy deposition inside the detector for the current event, to average over all detected hits, and weight the average by the relative energy deposit per hit, respectively.

To demonstrate the utility of having the full event topology available, \cref{fig:vertexPos} shows in colour scale the position of those \ce{^{234}Th} decays that deposit energy in the target crystal of the \texttt{TUM93A} detector module. For a given event stored in raw data, this is easily done: retrieve \texttt{TUM93A} from the \texttt{Detectors} collection, check if its \texttt{Hits} collection is non-empty, and in this case access the \texttt{Position} stored in the \texttt{StartState} of the \texttt{PrimaryTrack}. Calculating the distance between the start vertex of the primary particle and the position of \texttt{AverageHit}, shows that \ce{^{234}Th} is a contaminant with short-range  decay radiation: over \qty{80}{\percent} of those decays that are visible in the detector are located within \qty{3}{\cm} around the detector.

Another example is shown in \cref{fig:decay234}: the decomposition of the energy deposition due to intrinsic \ce{^{234}Th} contamination into distinct de-excitation pathways by identifying the intermediate nuclear levels. Because the full event topology is stored, a simple ROOT macro can recursively transverse the topology and identify the first and last level in the decay cascade, and partition the data accordingly (for more details see \cref{sec:detectorresponse:modelling,sec:detectorresponse:data}).

\subsection{Decoupling of Simulation and Data Processing}
\label{sec:datalayout:decoupling}
\begin{figure*}
\centering
\includegraphics{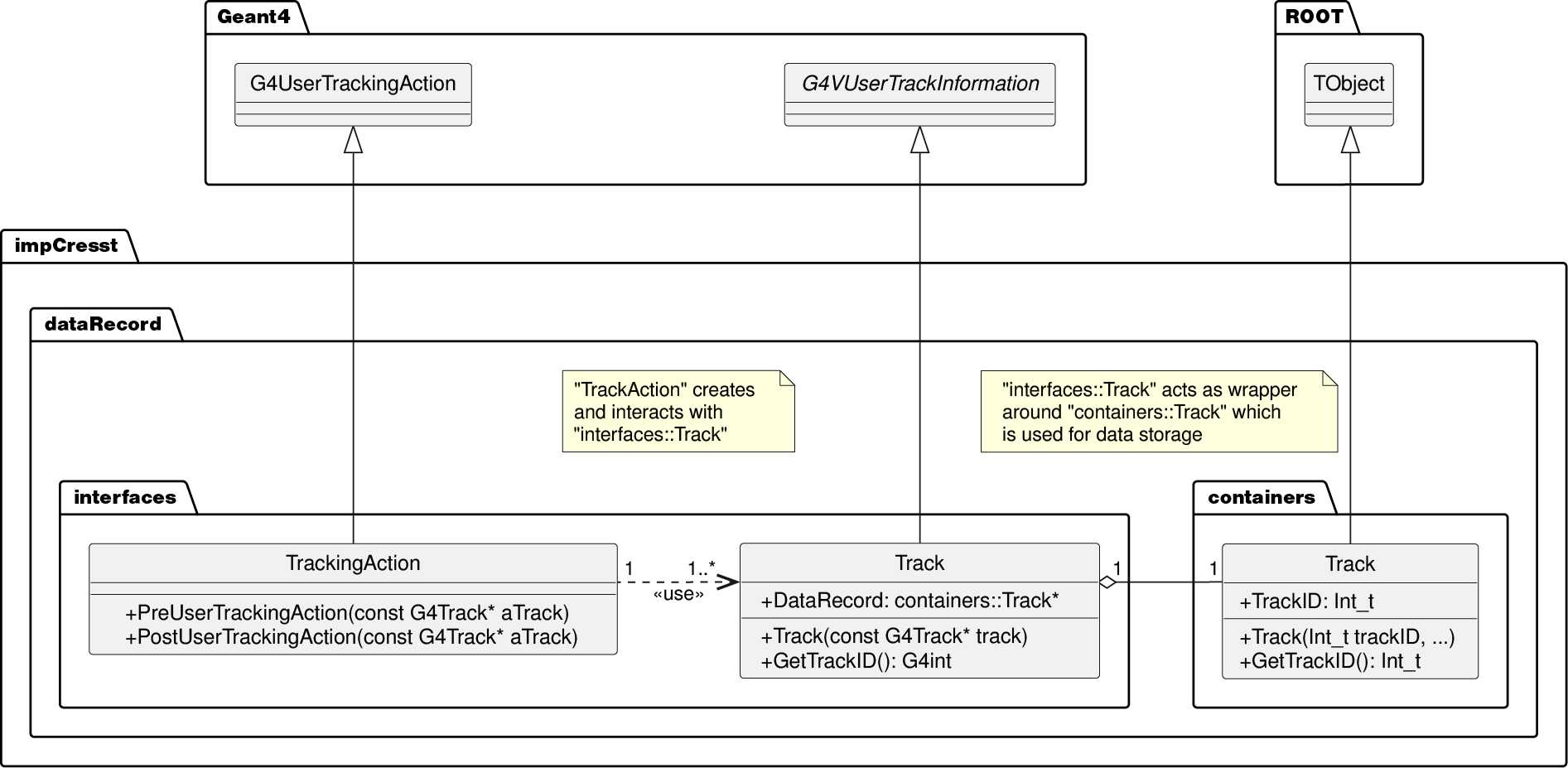}
\caption{\label{fig:dataBridge}Simplified UML class diagram of the bridge design pattern used to separate dependencies on \geant{} and ROOT using the track classes as example. For details, see text.}
\end{figure*}

To decouple the interface of our data container from any dependencies on \geant{}, we apply the \emph{bridge} pattern \cite{Gamma:2011}. It is illustrated in \cref{fig:dataBridge} using the track classes as example: an object of type \texttt{interfaces::Track} is created and called by the user specialization of the \geant{} user action class \texttt{G4UserTrackingAction}. It provides all functionalities to extract data from a \texttt{G4Track} object. Consequently, it has dependencies on \geant{} and exposed \geant{} data types, like \texttt{G4int}, on its interface. Internally, it instantiates an object of type \texttt{containers::Track} to store the extracted data once they are converted to ROOT data types, e.g., \texttt{G4int} to \texttt{Int\_t}. This container class has only dependencies on ROOT and is passed on to \texttt{containers::Event} to be stored in the output ROOT file. Hence, once the simulated data are stored, they can be further processed or analysed without any \geant{} dependencies.

The data structure of the processed files depends on the actual processing. See \cref{fig:processedFileFormat} in \cref{sec:detectorresponse} for the data structure in case of processing with \cresstds{}.

\subsection{File Size Reduction}
\label{sec:datalayout:size}

The resulting ROOT file is compressed with the \texttt{kLZMA} algorithm, as it is more efficient than ROOT's default \texttt{kZLIB} \cite{Zhang:2017,code:root:RCompressionSetting:EAlgorithm,code:root:TFile}. We chose a compression level of 1, as it balances the gained compression efficiency (\qty{-28}{\percent}) with the run time penalty (\qty{+16}{\percent}). Higher compression levels, like ROOT's recommended value of 7 for \texttt{kLZMA} or the maximum value of 9 \cite{code:root:RCompressionSetting,code:root:RCompressionSetting:ELevel}, result in only minimal improvements but significant penalties, see \cref{tab:compression}. For easier handling, the file is automatically split every \qty{50}{\giga\byte}.
\begin{table}[t]
\caption{Relative size of the \impcresst{} raw data file and relative run time of the simulation for different compression settings of ROOT, consisting of compression algorithm and compression level. Values are normalized to ROOT's standard settings, indicated with an asterisk (\textasteriskcentered). As test case, \num{1e3} decays of \ce{^{234}Th} in the \cawo{} target crystal of \texttt{TUM93A} were simulated. In this test case, \qty{100}{\percent} corresponds to an absolute file size of \qty{215}{\mega\byte} and an absolute run time of \qty{94}{\second} on the used computer. To avoid statistical fluctuations on the file size, each setting was simulated with the same random number generator seed. The spread of the run time corresponds to a relative uncertainty of \qty{7}{\percent}.}
\label{tab:compression}
\begin{tabular*}{\linewidth}{@{\extracolsep{\fill}} c l S[table-format=3] S[table-format=3]}
\toprule
\multicolumn{2}{c}{Compression setting} & {File size} & {Run time}\\
\multicolumn{2}{c}{Algorithm, level} & {/ \unit{\percent}} & {/ \unit{\percent}}\\\midrule
&\texttt{kLZ4, 1} & 113 & 97 \\
\textasteriskcentered & \texttt{kZLIB,1 } & 100 & 100 \\
&\texttt{kOldCompressionAlgo, 1} & 100 & 90 \\
&\texttt{kZSTD, 1} & 80 & 99 \\
&\texttt{kLZMA, 1} & 72 & 116\\
&\texttt{kLZMA, 7} & 68 & 249\\
&\texttt{kLZMA, 9} & 68 & 258\\
\bottomrule
\end{tabular*}
\end{table}

To control the file size in exchange for less detailed data storage, macro commands are provided:
\begin{itemize}
    \item To not store \texttt{AverageHit} and \texttt{EnergyWeighted Hit}, see \cref{lst:exampleMacro}, lines~\ref{lst:macro:storeAverageHit},\ref{lst:macro:storeEnergyWeightedHit}.
    \item To not store all tracks per event, but only those that cause a hit, see \cref{lst:exampleMacro}, line~\ref{lst:macro:storeAllTracks}. This setting is suitable, e.g., if only the particles that deposit energy are of interest. However, if one wants to backtrack where these particles originate, one needs to store all tracks to keep the whole chain of parent-child tracks.
    \item To not store hits, see \cref{lst:exampleMacro}, line~\ref{lst:macro:storeHits}. This setting is suitable, e.g.,  if only the sum energy deposition is of interest, or the cumulative values per event of user defined scorers. In this case, a post-processing with \cresstds{} is not possible any more.
    \item Only store the primary particle, see \cref{lst:exampleMacro}, line~\ref{lst:macro:storePrimaryTracks}. This setting is suitable, e.g., for studying the transmission through a radiation shield.
    \item To store only a reduced data set (\texttt{EnergyDeposit}, \texttt{VisibleEnergyDeposit}, \texttt{HitTime} and \texttt{StepLength}) per hit, see \cref{lst:exampleMacro}, line~\ref{lst:macro:storeFullHitData}.
\end{itemize}
\Cref{tab:dataDetail} illustrates the impact of these settings on the file size using the example of \num{1e3} decays of \ce{^{234}Th} as intrinsic contamination of the \cawo{} target crystal of \texttt{TUM93A}. As expected in this use case, the largest size reduction, without losing the opportunity to process the data with \cresstds{}, can be achieved by storing only  the reduced data set per hit. Therefore, this is \impcresst{}'s default setting. Switching the storage of any physics information off (eighth row from top), reveals the relatively small overhead of the data structure itself.

In case of \emph{intrinsic} background, switching off \texttt{storeAllTracks} does not reduce the file size significantly, as each track usually causes at least one hit. Hence, it is stored anyway because \texttt{storeHitTracks} is switched on by default. Contrary, for simulation of \emph{extrinsic} background, \texttt{storeAllTracks=false} is an efficient way to reduce the file size.

\begin{table}[thb]
\caption{Relative size of the \impcresst{} raw data file for different data configurations controlled by the given macro commands. A bullet (\textbullet) indicates the macro command is set to \texttt{true}, an open bullet (\textopenbullet) indicates it is set to \texttt{false}. The asterisk (\textasteriskcentered) indicates the default configuration. As test case, \num{1e3} decays of \ce{^{234}Th} in the \cawo{} target crystal of \texttt{TUM93A} were simulated. In this test case, \qty{100}{\percent} corresponds to an absolute file size of \qty{152}{\mega\byte}. To avoid statistical fluctuations on the file size, each setting was simulated with the same random number generator seed.}
\label{tab:dataDetail}
\begin{tabular*}{\linewidth}{@{\extracolsep{\fill}} c c c c c c c c S[table-format=3.2]}
\toprule
\multicolumn{8}{c}{Macro commands} & {Relative size / \unit{\percent}}\\ \cmidrule{1-7}
&\rotatebox{90}{\texttt{storeAverageHit}} & \rotatebox{90}{\texttt{storeEnergyWeightedHit}} & \rotatebox{90}{\texttt{storeAllTracks}} & \rotatebox{90}{\texttt{storeHitTracks}} & \rotatebox{90}{\texttt{storePrimaryTrack}} & \rotatebox{90}{\texttt{storeFullHitData}} & \rotatebox{90}{\texttt{storeHits}} & \\\midrule 
& \textbullet & \textbullet & \textbullet & \textbullet & \textbullet & \textbullet & \textbullet & 100.00\\
& \textopenbullet & \textbullet & \textbullet & \textbullet & \textbullet & \textbullet & \textbullet & 99.93\\
& \textopenbullet & \textopenbullet & \textbullet & \textbullet & \textbullet & \textbullet & \textbullet & 99.84\\
& \textopenbullet & \textopenbullet & \textopenbullet & \textbullet & \textbullet & \textbullet & \textbullet & 97.02\\
& \textopenbullet & \textopenbullet & \textopenbullet & \textopenbullet & \textbullet & \textbullet & \textbullet & 64.77\\
& \textopenbullet & \textopenbullet & \textopenbullet & \textopenbullet & \textopenbullet & \textbullet & \textbullet & 64.73\\
& \textopenbullet & \textopenbullet & \textopenbullet & \textopenbullet & \textopenbullet & \textopenbullet & \textbullet & 22.26\\
& \textopenbullet & \textopenbullet & \textopenbullet & \textopenbullet & \textopenbullet & \textopenbullet & \textopenbullet & 0.67\\
\textasteriskcentered & \textopenbullet & \textopenbullet & \textopenbullet & \textbullet & \textbullet & \textopenbullet & \textbullet & 54.96\\
& \textopenbullet & \textopenbullet & \textopenbullet & \textopenbullet & \textbullet & \textopenbullet & \textopenbullet & 0.71\\
\bottomrule
\end{tabular*}
\end{table}

\subsection{Metadata Annotation}
\label{sec:datalayout:metadata}
To enable a robust data provenance or data lineage, each created output file contains also an object of type \texttt{dataRecord::containers:: Record} which automatically logs certain metadata about the simulation. In case \impcresst{} was compiled using the provided CMake build configuration \cite{code:cmake} from a local Git repository \cite{code:git}, it will be aware of the used \impcresst{} version, the used Git repository, and if it contains unreleased modifications\footnote{During configuration of the build process, CMake will extract these data from the local Git repository and make it available to the build process as preprocessor definitions via its \texttt{add\_compile\_definitions} command \cite{code:cmake:addCompileDefinitions}. Within \impcresst{}, the functions within the name space \texttt{impCresst::buildInfo} access these preprocessor definitions, process them, and forward the data to the \texttt{containers::Recorder} object.}.  In the latter case, the related Git commit hash is given. Also, the used macro file to configure the \impcresst{} simulation, its start and end time, and the machine on which it was run are noted, together with other metadata cf.\ \cref{fig:rawFileFormat}. Also \cresstds{} adds automatically a similar record object, see \cref{fig:processedFileFormat}.

\section{Detector Response}
\label{sec:detectorresponse}
As mentioned in \cref{sec:workflow}, our simulation procedure consists of two steps. First, \impcresst{} simulates the microphysics (\cref{sec:physics}) and stores the raw data with all relevant information, as illustrated in \cref{fig:rawFileFormat}. The second step is post-processing the raw data with \cresstds{} to apply the real detector response effects of finite time and energy resolution, and scintillation light detection. That is, out of the raw hits of type \texttt{dataRecord::containers::Hit} recorded by a sensitive detector (\cref{sec:particleTracking}) and contained in a \texttt{Detector} object (\cref{fig:rawFileFormat}), it constructs what we call \emph{detector hits} of type \texttt{cresstDS::containers::DetectorHit}, which are comparable with the data measured with real \cresst{} detector modules (\cref{sec:detectorresponse:modelling}). The response models are applied by configurable data processors specific for each detector (\cref{sec:detectorresponse:processing}) and the processed data are stored again as ROOT files (\cref{sec:detectorresponse:data}). 

\subsection{Detector Response Modelling}
\label{sec:detectorresponse:modelling}
\begin{figure}
\centering
\includegraphics{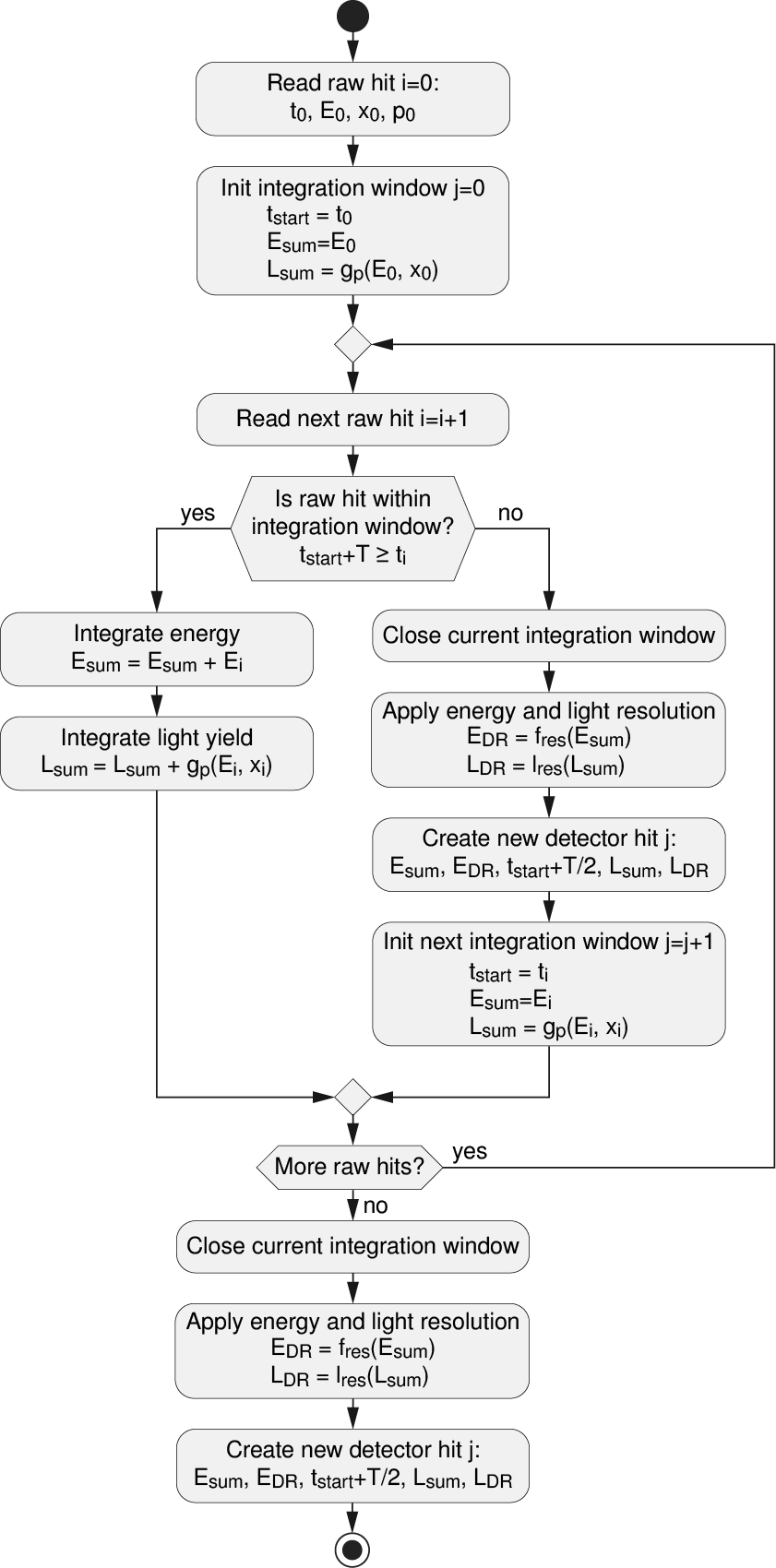}
\caption{\label{fig:detectorResponseMachine}UML activity diagram of the detector response  modelling: \emph{Detector hits} (objects of type \texttt{cresstDS::containers::DetectorHit}) are created by integrating \emph{raw hits} (objects of type \texttt{dataRecord::containers::Hit}), where the duration $T$ of the \emph{integration window} represents the time resolution of the detector. For each raw hit, the scintillation light yield is calculated from the deposited energy $E_i$ by the parametrization $g$, which may also depend on the step length $x_i$ and particle $p$ that causes the raw hit. After the integration window closes, the energy resolution is applied via a randomization of the integrated energy $E_\mathrm{sum}$ by an empirical parametrization $f_\mathrm{res}$. Similarly, the scintillation light yield $L$ is integrated and randomized by $l_\mathrm{res}$. The index ``DR'' indicates quantities with applied detector response.}
\end{figure}

The first effect of the detector response that is applied is the finite time resolution. We model it as the time period $T$ over which the raw hits are integrated with a finite-state machine consisting of two states: an open or closed integration window, see \cref{fig:detectorResponseMachine}. The procedure starts by reading the first raw hit which triggers the opening of an integration window. Then, the energy $E_i$ deposited by a \emph{raw} hit $i$ falling within this window is summed to $E_\mathrm{sum}$. Once the integration window closes, $E_\mathrm{sum}$ is stored in a new \emph{detector} hit and the next raw hit opens another integration window. We found that processing with an empirical value of $T=\qty{2}{\ms}$ results in data matching the real data taken with \cresst{} detector modules.

\begin{figure*}
    \centering
    \includegraphics{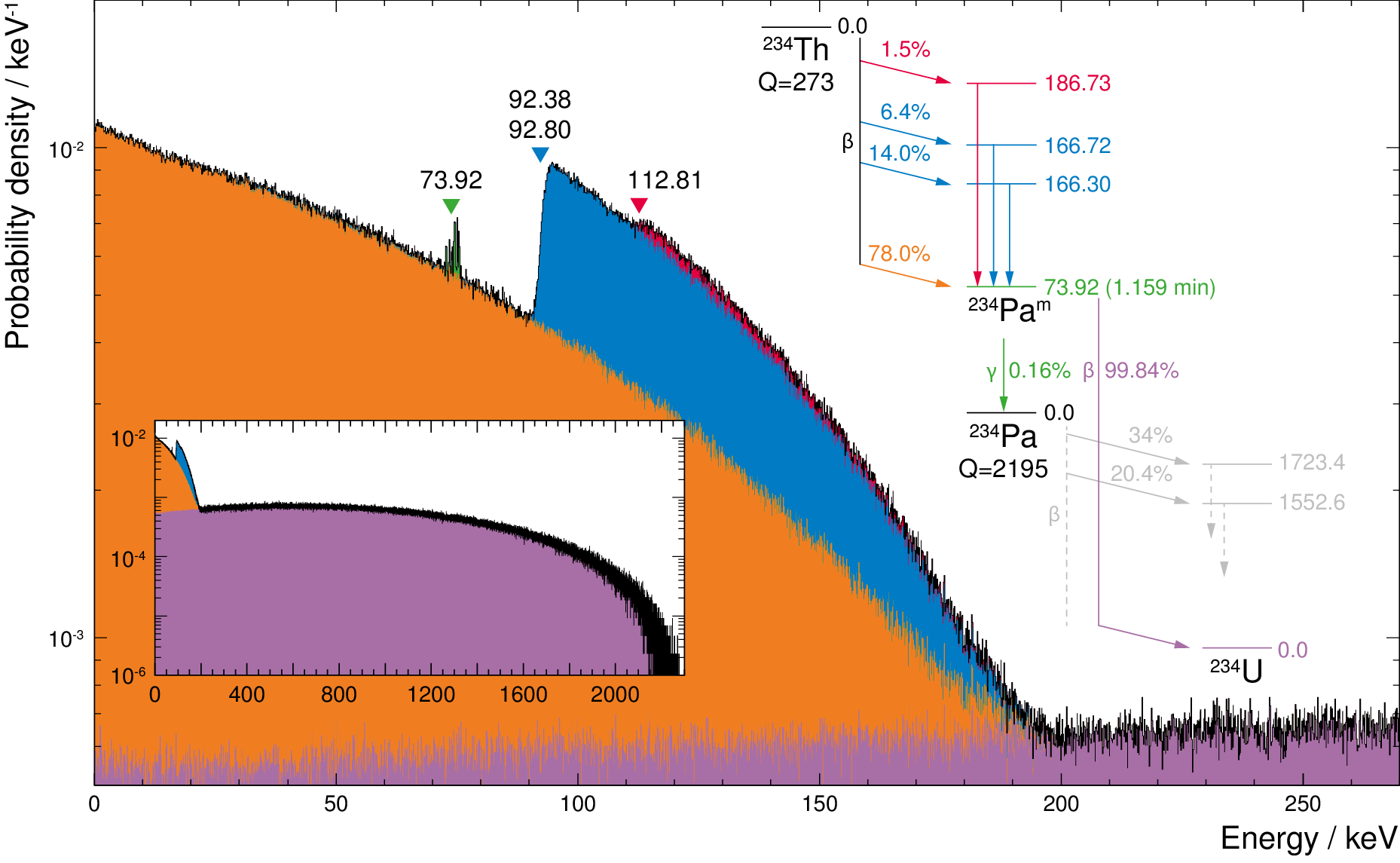}
    \caption{Simulation of \num{4.3775e6} decays of \ce{^{234}Th} within the \cawo{} target crystal of the TUM93A detector module with applied detector response model (bin size of \qty{100}{\eV}). The \emph{inset} zooms out to \qty{2.3}{\MeV} (bin size of \qty{1}{\keV}). Due to the isomeric state $\ce{^{234}Pa}^\mathrm{m}$ [\qty{73.92}{\keV}] the decay results in prompt (\emph{orange}, \emph{blue}, and \emph{red} filled histograms) and delayed (\emph{violet} filled histogram and \emph{green} peak at \qty{73.92}{\keV}) detector hits. The simulation ends once the ground state of one of the daughter nuclides \ce{^{234}Pa} and \ce{^{234}U} is reached. Hence, the decay of \ce{^{234}Pa} itself is not included and has to be simulated separately. Data for the simplified level scheme are taken from \cite{livechart} and given in keV. The decay scheme of \ce{^{234}Pa} is only partial given and implied via the \emph{dashed, grey} lines.}
    \label{fig:decay234}
\end{figure*}

The impact of the time resolution is evident in case of the \ce{^{234}Th} decay (see \cref{fig:decay234}\footnote{For the technicalities how the shown decay pathways were identified within a \impcresst{}/\cresstds{} based workflow, see also \cref{sec:dataformt:layout,sec:detectorresponse:data}.}): the decay cascade goes through the isomeric state $\ce{^{234}Pa}^\mathrm{m}$, whose half-life is with $T_{1/2}=\qty{1.159}{\minute}$ long compared to \cresst{}'s time resolution. Consequently, each \ce{^{234}Th} decay has a high probability to cause \emph{two} detector hits: a first prompt hit belonging to the \emph{red}, \emph{blue}, or \emph{orange} populations in \cref{fig:decay234}, and a delayed hit belonging to the \emph{green} or \emph{violet} populations. For all other excited levels of \ce{^{234}Pa} the nuclear de-excitation following the \textbeta{} decay of \ce{^{234}Th} is fast compared to the time resolution, hence the \textbeta{} spectra are shifted by the energy of the emitted \textgamma{}-ray (\emph{blue} and \emph{red} populations). Due to the way we split the nuclear decay chain (see \cref{sec:physics:radioactivity}), the decay of \ce{^{234}Pa} itself is not included in \cref{fig:decay234}, but has to be  simulated separately, see e.g., Ref.~\cite{Angloher:2024}. 

The second effect applied is the finite energy resolution. At the closed window-state, the integrated energy $E_\mathrm{sum}$ is randomized by a function $f_\mathrm{res}(E)$ that parametrizes the energy resolution and is of ROOT's \texttt{TF1} \cite{code:root:TF1} type. For \cresst{}, this is usually a Gaussian centred at $E_\mathrm{sum}$ with an energy-dependent standard deviation empirically obtained from (calibration) measurements. However, the user is free to specify any function supported by the \texttt{TF1} class. As \cresstds{} also saves $E_\mathrm{sum}$, i.e., the integrated energy deposition \emph{without} randomization, it allows applying a different energy resolution without needing the raw data files. Hence, an easy adaptation of the processed data to different analysis workflows of the real measurements, yielding different energy resolutions, is possible.

As \impcresst{} does not simulate the light emitted by \cresst{}’s scintillating target crystals by default (\cref{sec:physics:optics}), also the quenched scintillation light yield $L$ is calculated in a linear approximation during post-processing with \cresstds{}. Similar to the energy integration, a light yield $L_i$ for each raw hit is calculated and summed to $L_\mathrm{sum}$ during the open-window state. At the closed-window state, $L_\mathrm{sum}$ is randomized by a resolution function $l_\mathrm{res}$ and assigned to the created detector hit. \cresstds{} offers two methods to calculate $L_i$: one based on Birk’s theory of scintillation \cite{Birks:1951}, originally developed for organic scintillators, and one based on empirical parametrizations.

For Birk's approach,
\begin{equation*}
\begin{split}
    L_i = g(E_{\mathrm{hit},i},x_i) &= \int_x \frac{\mathrm{d}L_i}{\mathrm{d}x} \mathrm{d}x\\
    &=S \cdot \frac{E_{\mathrm{hit},i}}{1+k B \frac{\mathrm{d}E}{\mathrm{d}x}}
\end{split}
\end{equation*}
we approximate the specific energy loss $\mathrm{d}E/\mathrm{d}x$ by $E_{\mathrm{hit},i}$ over the related step length\footnote{We note, that, if Birk's approach is applied, a certain fine-tuning of the maximum step size may be needed to achieve satisfactory results, see \cite{Fuss:2017}. \impcresst{} provides for this the macro command on line~\ref{lst:macro:stepLimit} of \cref{lst:exampleMacro}.} $x$, and treat $S$ and $kB$ as empirical parameters.

Alternatively, an empirical parametrization of the quenched light yield $L_i=L_{i,p}=g_p(E_{\mathrm{hit},i})$ is used, where $p$ is the type of particle that caused the raw hit. In case of \cawo{}, we provide default values for Birk's approach from Ref.~\cite{Lang:2009} and parametrizations of the quenched light yield from Ref.~\cite{Strauss:2014}. Again, the user is able to enter any parametrization supported by the \texttt{TF1} class. For \cresst{}, we found good agreement between simulation and real data by using parametrized $L_{i,p}$, see \cite{Fuss:2017,Fuss:2022}.

The light resolution function $l_\mathrm{res}$ can be any function supported by the \texttt{TF1} class. Similar to the case of the energy resolution, for \cresst{} it is usually a Gaussian with energy-dependent standard deviation empirically obtained from (calibration) measurements.

\subsection{Configurable Data Processing}
\label{sec:detectorresponse:processing}
The \cresst{} experiment uses multiple detectors, so the processing with \cresstds{} has to be modular, flexible, and dynamic: it identifies all sensitive detectors present in the raw data and creates for each an object of type \texttt{DetectorHitBuilder}, which is responsible to apply the specific response of a particular detector and create \emph{detector} hits out of \emph{raw} hits. Each \texttt{DetectorHitBuilder} is registered with a \texttt{DataProcessor} object that reads in the raw data, identify the raw hits recorded by a given sensitive detector and dispatch them to the corresponding \texttt{DetectorHitBuilder}. The \texttt{DetectorHitBuilder} objects are configured via structured files parsed by the \textit{\texttt{libconfig}} library \cite{code:libconfig}, hereafter called \emph{detector configuration files}, see \cref{lst:exampleCfg} for an example.

This file consists of several \emph{groups} of \emph{settings}, each setting a key-value pair. Here, a group corresponds to one detector and  contains the detector response parameters and the detector name. For example, the group on lines \ref{lst:cfg:TUM93aStart} to \ref{lst:cfg:TUM93aEnd} of \cref{lst:exampleCfg} corresponds to the TUM93A detector module and the setting on line \ref{lst:cfg:keyValue} specifies the name. A group is matched to the related \texttt{DetectorHitBuilder} object by the sensitive detector name.

In case no matching group is found, the \emph{default} group (cf.\ \cref{lst:exampleCfg}, lines \ref{lst:cfg:defaultStart}-\ref{lst:cfg:defaultEnd}) is used and \cresstds{} issues a warning. The default group is also used to substitute settings which are missing in a detector specific group. Hence, common standard values can be specified once in the default group and do not have to be repeated for each detector. For example, for the TUM93A group in \cref{lst:exampleCfg} only the energy resolution function $f_\mathrm{res}$ is specified for this module (line~\ref{lst:cfg:energyResolution}), whereas the light resolution function $l_\mathrm{res}$ is taken from the default group (line~\ref{lst:cfg:lightResolution}). As libconfig files are human-readable, the user can easily provide their own groups, including the default group. 

\subsection{Processed Data Layout}
\label{sec:detectorresponse:data}
\begin{figure*}
\centering
\includegraphics{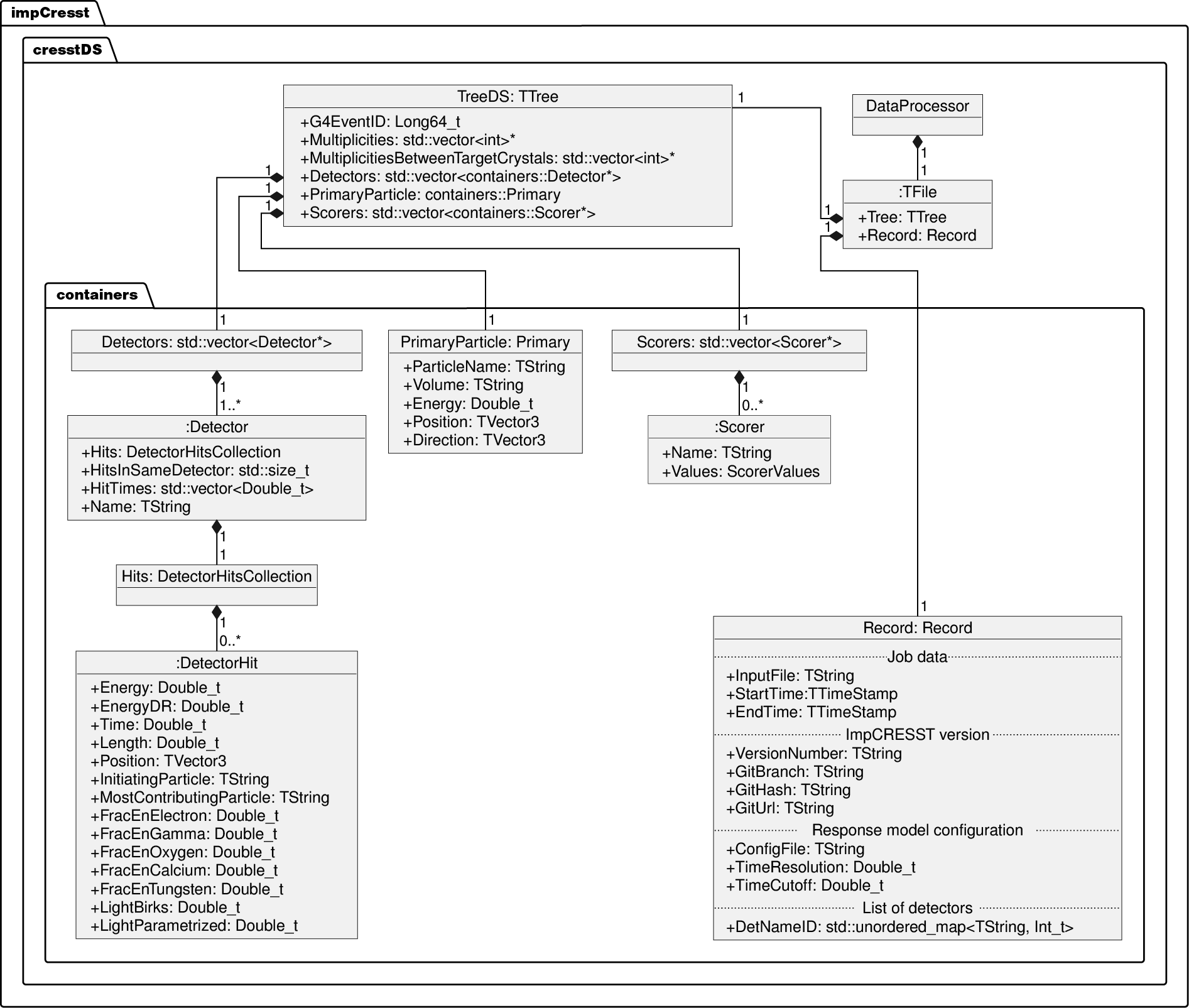}
\caption{\label{fig:processedFileFormat}Simplified UML object diagram of the data structure used by \cresstds{} to store the processed data in ROOT files. For the sake of simplicity, not all data members of \texttt{Hit} and \texttt{Record} are shown. For details see text.}
\end{figure*}

For high level analysis of the processed data, it is also essential to store further data in addition to the recorded detector hits, e.g., the primary particle, the total number of simulated events, and the metadata record. The \cresstds{} data layout includes all these data and is shown in \cref{fig:processedFileFormat}. Similar to the raw data (\cref{fig:rawFileFormat}), also the processed data per event are stored as leaves on a ROOT \texttt{TTree}. However, as the linkage between particle tracks and raw hits is lost during integration, there is no need for a monolithic event class any more. Instead, we use several \texttt{TBranches} \cite{code:root:TBranch}: the branch \texttt{PrimaryParticle}  stores data of type \texttt{cresstDS::containers::Primary} related to the primary particle from the \impcresst{} simulation. Similar to the raw data layout (\cref{fig:rawFileFormat}), the collection of detector hits is stored per detector object of type \texttt{cresstDS::containers::Detector}; a collection of detector objects is stored as branch \texttt{Detectors}. Data from the scorers (\cref{sec:particleTracking}) are stored under the branch \texttt{Scorers}.

Due to the accumulation of raw hits to detector hits, the size of the processed files is usually significantly smaller than that of the raw file, making it suitable for long-term data storage. As an example, \cref{fig:decay234} is based on \qty{683}{\giga\byte} of raw data reduced by a factor of \num{197} to \qty{3.46}{\giga\byte} of processed data.

Albeit the linkage between raw hits and tracks is lost during processing with \cresstds{}, it is still possible to analyse the resulting detector hits with respect to raw track properties, as illustrated by \cref{fig:decay234}: First, the raw data get partition accordingly to the criterion under investigation (here: nuclear decay pathways). Afterwards, each data partition is processed on its own. This approach can be conveniently realised within a workflow managed by, e.g., nextflow (see \cref{sec:environment}).

\section{Traceability and Reproducibility}
\label{sec:environment}
\impcresst{} is designed with a strong emphasis on traceability and reproducibility, ensuring that every simulation performed with the package can be reliably replicated and independently verified.

The metadata tagging of the simulation files (see \cref{sec:datalayout:metadata}) allows identifying the \impcresst{} version which was used for each simulated data set as well as the used macro file. If the raw data were processed with \cresstds{}, also the used detector configuration file is noted. We adhere to semantic versioning principles \cite{semver}, where each release is clearly labelled to reflect the nature of the changes — major, minor, or patch updates -- and uniquely identifiable.
The releases and the complete codebase are managed under Git-based version control \cite{code:git}, providing a transparent record of all modifications. Furthermore, each release contains a comprehensive set of installation instructions strictly specifying the dependencies of \impcresst{} and its computational environment\footnote{For \impcresst{} version 8.0.0, the environment is based on AlmaLinux version 9.5 \cite{code:almalinux}, GCC version 8.3 \cite{code:gcc}, \geant{} version 10.6.3 \cite{code:geant}, and ROOT version 6.32.08 \cite{code:root}; following the C++17 standard. For more details, see the files \texttt{./README.md}, \texttt{./CONTRIBUTING.md}, and \texttt{./doc/codingStyle.md} distributed with the \impcresst{} package.}. This specification and the versioning enable users to reproduce the computational environment precisely as used in published analyses with the software. Running the same macro file using the same RNG seed in the same environment will result in the same simulation data.

To enhance reproducibility across platforms, we containerize the environment using container images that encapsulate a versioned \impcresst{} release and its dependencies. The container image is created via a \emph{continuous integration} and \emph{continuous deployment} (CI/CD) pipeline integrated with our internal Git repository. We employ a nested containerization strategy: A dedicated, internal Git repository hosts the CI/CD pipeline to build a container for \textsc{Geant4} and ROOT\footnote{We decided against using the official \geant{} and ROOT containers, to ensure consistent build settings across \geant{}, ROOT, and \impcresst{}.}, which serves as the base environment for building the \impcresst{} container\footnote{The recipe for a build process is given in the format of a Dockerfile. For user convenience, the recipes for both the \textsc{Geant4}/ROOT and \impcresst{} build processes are distributed with the \impcresst{} package: \texttt{./examples/geant4RootContainer/Dockerfile} and \texttt{./Dockerfile}, respectively.}. As the \geant{} and ROOT environment usually stays stable over several \impcresst{} releases, this nesting avoids superfluous and time-consuming recompilation of \geant{} and ROOT for each new \impcresst{} release. The CI/CD pipeline is triggered when any branch on our internal \impcresst{} Git repository is ``tagged'', e.g., for a new release of \impcresst{}. The newly created container is automatically deployed to a collaboration internal container registry. Authorized users can pull the container easily, either as a Docker \cite{code:docker} or Apptainer\footnote{Formerly known as Singularity.} \cite{code:apptainer} image. The latter is the preferred container framework for deployment in HPC environments. The use of a container image ensures that simulations performed with identical containers will operate in bitwise-identical environments regardless of the underlying operating system or hardware setup, a critical aspect for scientific reproducibility.

The containerisation also facilitates the lazy parallelism we use for production runs (see \cref{sec:workflow}) at \emph{The Max Planck Computing and Data Facility} (MPCDF, Munich, \cite{MPCDF}) and \emph{The Cloud Infrastructure Platform} (CLIP, Vienna, \cite{CLIP}). Orchestrating scalable parallel computation by launching numerous independent simulation instances in parallel can be either done by simple shell scripts or with modern scientific workflow managers like nextflow \cite{code:nextflow}\footnote{We provide minimal working examples for both bash and nextflow based simulation workflows within the released \impcresst{} package, see \texttt{./examples/simulationWorkflowBash/} and \texttt{./examples/simulationWorkflowNextflow/}, respectively.}. In \cresst{}, we have good experiences with the latter \cite{kluck:2026}: it allows to easily script the workflow (several parallel runs of \impcresst{}, after each run post-processing the raw data with \cresstds{}, finally the merging of the processed data), it transparently loads the container image, and handles the communication with the job scheduler, in our case Slurm \cite{code:slurm}. It also enhances reproducibility, as not only the simulation itself can be reproduced but also the whole simulation workflow by rerunning the nextflow script.

\section{Conclusion and Outlook}
\label{sec:conclusion}

With this article, we presented \impcresst{}, an open source, \geant{}-based Monte Carlo simulation tool for cryogenic, solid-state particle detectors developed for the \cresst{} experiment searching for DM. It is adaptable to other rare event searches using solid-state detectors without difficulty.

To our knowledge, \impcresst{} is the only publicly available simulation program that is specially tuned for: a heterogeneous and fast evolving detector environment by offering a flexible geometry handling for rapid prototyping of new detectors via direct read-in of CAD files (see \cref{sec:geometry}); background studies at the keV range, providing detailed modelling of electromagnetic interactions and nuclear de-excitations (\cref{sec:physics}); versatile and easy-to-use particle generators for radiogenic and cosmogenic background sources. \impcresst{} includes also the \texttt{ContaminantSource}, a particle generator for the easy ``contamination'' of a flexible geometry selection with radioactive bulk or surface contaminants.  (\cref{sec:particleGenerators}). To support a fast experimental development process, \impcresst{} features a 
data persistency of the whole event topology as base for developing tuned data post-processing and analysis. Automatic metadata annotation enables a robust data provenance (\cref{sec:dataformat}). For standard data analysis, we provide the auxiliary \cresstds{} tool to flexibly apply parametrized models of time and energy resolution. Its detector specific parameters are stored in an easy-to-maintain set of human-readable configuration files (\cref{sec:detectorresponse}). \impcresst{} is well suited to be deployed in an HPC environment in a traceable and reproducible way. As an example we described a workflow based on Apptainer for containerization, Slurm for job scheduling, and nextflow for workflow management in \cref{sec:environment}.

Also simulations of liquid and gaseous detector targets, and of interactions above the \qty{10}{\MeV} scale, are technically possible with \impcresst{}; however, we conduct no validation or tuning as these use cases are out of scope for \cresst{}. Similar, we do not support the modelling of electron-hole interactions in semiconductor detectors. 

For the future, several enhancements to \impcresst{} are planned: To speed-up time-consuming studies of the shielding against the extrinsic, ambient \textgamma{} background, we are currently implementing the biasing scheme specified in Ref.~\cite{Zatschler:2025}. To improve the fidelity of simulating radioactive contaminants close to the surface, a performance optimized approach to place contaminants in a detailed surface geometry based on measurements of real-world roughness profiles \cite{Gruener:2024,Gruener:2025} was developed in our group and is available at \cite{code:SCoRe4}; it will be included in future releases of \impcresst{}. Those roughness-profiles could also be used to improve our optical simulation capabilities. For electromagnetic interactions, we plan to switch the default EMPC to \texttt{G4EmLivermore} as we found in Ref.~\cite{Kluck:2025} that it is more robust against parameter variation than the currently used \texttt{G4EmStandardPhysics\_option4} for our use case. To improve the modelling of background processes that are currently sub-dominant for \cresst{}, but may become relevant at lower detection thresholds in the future, we are looking into including coherent photon-nucleus scattering \cite{Omer2017,Omer2018} and coherent neutron scattering \cite{Cai:2019}. We may also include phonon interaction by interfacing with G4CMP \cite{Kelsey:2023}. Generally, we aim for the near future to upgrade to the recent \geant{} version 11 to profit especially from the improved precision in neutron physics \cite{Zmeskal:2023,Zmeskal:2024} and modelling of the nuclear de-excitation \cite{Mendoza:2020}. Consequently, \impcresst{} will also feature the default multithreading of \geant{} version 11. From the technical side, we are considering implementing automatic tests of well-defined physics cases to prevent regressions during the software development. We plan also to offer the \texttt{ContaminantSource} as a stand-alone software library that can be easily integrated with other \geant{}-based applications.

As the usage by the \textsc{Nucleus} and \textsc{Cosinus} collaborations proved, \impcresst{} is a well suited tool for other experiments dealing with cryogenic, solid-state particle detectors in a heterogeneous environment. 
Hence, starting with its current version 8.0.0 \cite{code:impcresst}, we release \impcresst{} as open-source code under the GPLv3 licence \cite{gpl}. A public mirror of our releases can be found at: \url{https://github.com/cresst-experiment/impcresst/}.

\backmatter

\bmhead{Acknowledgment}

This work has been funded in part by the Deutsche Forschungsgemeinschaft (DFG, German Research Foundation) under Germany’s Excellence Strategy – EXC2094 – 390783311 and through the Sonderforschungsbereich (Collaborative Research Center) SFB1258 ``Neutrinos and Dark Matter in Astro- and Particle Physics'', by the BMBF 05A23WO4 and 05A23VTA, by the Austrian Science Fund (FWF) \href{http://dx.doi.org/10.55776/PAT1239524}{DOI:10.55776/PAT1239524} and by \href{http://dx.doi.org/10.55776/I5420}{DOI:10.55776/I5420}. J.~Burkhart and H.~Kluck were funded through the FWF project ``ELOISE'' \href{http://dx.doi.org/10.55776/P34778}{DOI:10.55776/P34778}. The Bratislava group acknowledges a partial support provided by the Slovak Research and Development Agency (project APVV-21-0377).

We gratefully acknowledge contributions to the \impcresst{} package from S.~Arora, C.~Fritz, E.~Gutmann, L.~Heckmann, A.~Karl, M.~Kirchner, V.~Palu\v{s}ová, L.~Pfaffenbichler, M.~Poneder, R.~Prinz, A.~Rabensteiner, F.~Stummer, C.~Türko\u{g}lu, and M.~Weiskopf.

A special acknowledgement is due to S.~Scholl, whose conception of the original code and the name ``\impcresst{}'' laid the foundation for this project. 

\bmhead{Declarations}
\begin{itemize}

\item \textbf{Conflict of interest/Competing interests:} The authors have no relevant financial or non-financial interests to disclose.
\item \textbf{Code availability:} The \impcresst{} package is available under: \url{https://github.com/cresst-experiment/impcresst/}

\item \textbf{Author contribution:} H.~Kluck conceptualized the paper and coordinated its writing. S.~Banik, R.~Breier, H.~Kluck, and V.~Mokina wrote the manuscript. Visualisations were done by J.~Burkhart and the remaining figures by H.~Kluck. The overall development of ``\impcresst{}'' has been carried out collaboratively by its steering committee (S.~Banik, R.~Breier, H.~Kluck and V.~Mokina) which is chaired by the maintainer (H.~Kluck). The sub package ``ContaminantSource'' was mainly developed by A.~Fuss. All authors reviewed and agreed to the article.
\end{itemize}

\begin{appendices}
\begin{onecolumn}
\section{Example Files}
\label{sec:appendix}
\Cref{lst:exampleMacro} provides an excerpt of the \impcresst{} macro file used for the simulation resulting  in \cref{fig:vertexPos}. A reduced version of the \cresstds{} configuration file used for \cref{fig:decay234} is provided as \Cref{lst:exampleCfg}.

To illustrate the structure of a \cresstds{} configuration file, \cref{lst:exampleCfg} shows a group of settings for the detector module TUM93A. However, at the time of publication, \cresst{}'s data taking run 37 is still ongoing, and hence no detector response parameters for TUM93A in this particular run are available yet. Therefore, the given energy resolution for TUM93A repeats the default values. For the \texttt{lightQuenchingParametrisation} setting, oxygen and tungsten are encoded as \texttt{O1} and \texttt{W1}, respectively; this will change in future versions of \cresstds{}.
\begin{lstlisting}[caption={An excerpt of an \impcresst{} macro file demonstrating how to: simulate \ce{^{234}Th} decays inside the copper parts of \cresst{}'s \texttt{Carousel37}, load detector modules from CAD files, create scorers in the parallel world, and configure the geometry, physics list, and data storage. Lines starting with a hash sign (\texttt{\#}) are comments. In the released \impcresst{} package, the complete file can be found as \texttt{./examples/contaminantRange234Th/range.mac}.}, label={lst:exampleMacro}, numbers=left]
########################################################################
# PreInit state

# Store RNG status to file and
# provide status also to user code
/run/storeRndmStatToEvent 1

# No detailed printout during the simulation run
/tracking/verbose 0

# The macro path is set via the environment variable
# IMPCRESST_MACRO_PATH, which is automatically set if the
# ./bin/thisImpCRESST.sh or ./bin/thisImpCRESST.csh script is sourced.
/control/getEnv IMPCRESST_MACRO_PATH
/control/macroPath {IMPCRESST_MACRO_PATH}

# Optional, activate optical physics
# /physics/activateOpticalPhysics true~\label{lst:macro:optics}~

# Optional, select a Electromagnetic Physics Constructor
/physics/emConstructor G4EmLivermore~\label{lst:macro:empc}~

########################################################################
# Init state

# Init Geant4's run manager, this is necessary before the geometry can
# be created
/run/initialize

# Geometry settings----------------------------------------------------

# Set check for overlaps and verbose level
/geometry/checkForOverlaps true
/geometry/verboseLevel 2

# Set setup to CRESST's full setup at LNGS
/geometry/setSetup CresstAtLNGS ~\label{lst:macro:cresst}~
# Select detector carousel of data taking run 37
/geometry/CresstAtLNGS/chooseCarousel Carousel37 ~\label{lst:macro:carousel}~

# Add detectors to carousel:
# /geometry/<Carousel>/addDetector <DetectorType> <NameOfDetector> <DetectorSlotPosition> <pathTo.ObjFile> <pathTo.macFile>
# In a default installation, the 'geometry' directory that
# contains the CAD drawings as OBJ files is located at
/control/alias IMPCRESST_GEOMETRY_PATH {IMPCRESST_MACRO_PATH}/../geometry

# CRESST3Detector
/geometry/Carousel37/addDetector CRESST3Detector TUM93A TUM93A {IMPCRESST_GEOMETRY_PATH}/CRESST3Detector/CRESST3.obj {IMPCRESST_GEOMETRY_PATH}/CRESST3Detector/CRESST3Detector.mac
# The remaining detector modules are similarily loaded from OBJ files,
# see the complete macro file for details.

# Tuning the scintillation quenching in optical simulations may require
# limiting the maximum step length in <material> to <stepLimit> in mm
# /geometry/setSteplimitInMaterial <material> <stepLimit> ~\label{lst:macro:stepLimit}~

# Build the setup
/geometry/buildSetup~\label{lst:macro:build}~

# Open the shieldign by moving right, left, and front part by 1 m
/geometry/CresstAtLNGS/ShiftShieldingRight 1.0 m~\label{lst:macro:shift}~
/geometry/CresstAtLNGS/ShiftShieldingLeft 1.0 m 
/geometry/CresstAtLNGS/ShiftShieldingFront 1.0 m

# Add scorer to the parallel world: score fluxes and energies of
# electrons and gammas that are incident on the CaWO4 target crystal 
# of the TUM93A detector module
/geometry/addScorer TUM93A_CaWO4 e- flux 0 in~\label{lst:macro:scorerFirst}~
/geometry/addScorer TUM93A_CaWO4 gamma flux 0 in
/geometry/addScorer TUM93A_CaWO4 e- energy 0 in
/geometry/addScorer TUM93A_CaWO4 gamma energy 0 in~\label{lst:macro:scorerEnd}~

# Build the parallel world
/geometry/buildParallelWorld~\label{lst:macro:parallelWorld}~

# Optional, enable visualization
# /control/execute ./mac/visualization/vis.mac

# Primary particle generation------------------------------------------

# Generate primary particle via the contaminant source
/source/type contaminantSource~\label{lst:macro:cosoSelect}~
/contaminantSource/verbose 0

# Use 234Th as contaminant
/gps/particle ion~\label{lst:macro:gpsFirst}~
/gps/ion 90 234
/gps/ang/type iso
/gps/energy 0 MeV~\label{lst:macro:gpsLast}~

# Confine contaminant to all copper parts within Carousel37
/contaminantSource/confineToMaterialInVolume CRESST_Cu Carousel37~\label{lst:macro:cosoConfine}~
/contaminantSource/init~\label{lst:macro:cosoInit}~

# Stop decay chain once decay of 234Th reaches a ground state
/grdm/nucleusLimits 234 234 90 90~\label{lst:macro:grdm}~

# Enable low-energy physics.
/process/em/fluo true~\label{lst:macro:lowEnergyStart}~
/process/em/auger true
/process/em/augerCascade true
/process/em/pixe true
/process/em/deexcitationIgnoreCut true~\label{lst:macro:lowEnergyEnd}~

# I/O settings--------------------------------------------------------

# Set name of output file
/data/setPrefix ./range_234Th_

# Select which data to store:
# Store all tracks, including those which don't create hits
/data/storeAllTracks true~\label{lst:macro:storeAllTracks}~
# Do store avereaged hits
/data/storeAverageHit true~\label{lst:macro:storeAverageHit}~
#/data/storeEnergyWeightedHit true~\label{lst:macro:storeEnergyWeightedHit}~
# Do not store hits at all
#/data/storeHits false~\label{lst:macro:storeHits}~
# Do not store track that created hits
#/data/storeHitTracks false~\label{lst:macro:storeHitTracks}~
# Do not store the primary track
#/data/storePrimaryTrack false~\label{lst:macro:storePrimaryTracks}~
# Do store the full data set per hit
#/data/storeFullHitData true~\label{lst:macro:storeFullHitData}~

# Start simulation-----------------------------------------------------

# Optional, print progress every 1000 events
/run/printProgress 1000

# Optional, re-run previous simulation by reading RNG status from file
#/random/resetEngineFrom ./seed.rndm~\label{lst:macro:reseed}~

# Simulate 1e6 events
/run/beamOn 1000000

\end{lstlisting}

\begin{lstlisting}[caption={An excerpt of a detector configuration file for \cresstds{}: it provides a group of default settings and the detector inventory, which consists of only one group of settings for the \texttt{TUM93A} detector module. In the released \impcresst{} package, the complete file can be found as \texttt{./examples/decayPaths234Th/detectors.cfg}.}, label={lst:exampleCfg}, numbers=left]
default = {~\label{lst:cfg:defaultStart}~
	name = "default";
	threshold = 0.0301;
	deadTime = 0;
	timeResolution = 2E6;
	energyResolutionIsGaussian = true;
	energyResolutionSigma = "std::sqrt(0.0045*0.0045 + 0.0104298*0.0104298 * (x*x - 0.0301*0.0301))";
	energyResolutionFunction = "";
	lightResolutionIsGaussian = true;
	lightResolutionSigma = "std::sqrt(0.094*0.094 + (0.185898 * x) + (0.0260744 * x * x))";~\label{lst:cfg:lightResolution}~
	lightResolutionFunction = "";
	lightBirksConstant_kB = 0.0185;
	lightBirksConstant_A = 1.;
	lightQuenchingParametrisation = ( 
		{
			particle = "O1";
			quenching = "0.07908 * (1. + 0.7088 * exp(-x*1000./567.1)) * x";
		},
		{
			particle = "Ca";
			quenching = "0.05949 * (1. + 0.1887 * exp(-x*1000./801.3)) * x";
		},
		{
			particle = "W1";
			quenching = "0.0208 * x";
		},
		{
			particle = "e-";
			quenching = "(0.938+4.6e-5*x*1000.) * (1-0.389*exp(-x*1000./19.34)) * x";
		}
	);
}{~\label{lst:cfg:defaultEnd}~
inventory = {detector =  (
        {{~\label{lst:cfg:TUM93aStart}~
            name  = "TUM93A_CaWO4_Bolometer";~\label{lst:cfg:keyValue}~
            threshold =  0.0515008;
            deadTime = 0;
            energyResolutionIsGaussian = true;
            energyResolutionSigma  = "std::sqrt(0.0045*0.0045 + 0.0104298*0.0104298 * (x*x - 0.0301*0.0301))";~\label{lst:cfg:energyResolution}~
        }~\label{lst:cfg:TUM93aEnd}~
    )}
\end{lstlisting}
\end{onecolumn}
\twocolumn

\end{appendices}
\newpage
\bibliography{main} 

\end{document}